\documentclass[iop]{emulateapj}

\usepackage{graphicx}
\usepackage{color}
\usepackage{amsmath,amssymb}
\usepackage{type1cm}
\usepackage{ulem}
\usepackage{mathrsfs}
\usepackage{lineno} 

\usepackage{xfrac}
\usepackage{hyperref}

\definecolor{DCW}{rgb}{0.5,0,1} 
\definecolor{lang}{rgb}{1,0,0} 

\def\ga{\,\,\raise0.14em\hbox{$>$}\kern-0.76em\lower0.28em\hbox{$\sim$}\,\,}
\def\la{\,\,\raise0.14em\hbox{$<$}\kern-0.76em\lower0.28em\hbox{$\sim$}\,\,}

\slugcomment{draft version \today, }

\shorttitle{Numerical simulation of photospheric emission in LGRBs}                                                      
\shortauthors{Ito et al.}

\begin{document}

\title{Numerical simulation of photospheric emission in long gamma-ray bursts: prompt correlations, spectral shapes, and polarizations}

\author{Hirotaka Ito\altaffilmark{1,2},  Jin Matsumoto\altaffilmark{3}, Shigehiro Nagataki\altaffilmark{2,4,5},  Donald C. Warren\altaffilmark{6},  Maxim V. Barkov\altaffilmark{1,7,8} and Daisuke Yonetoku\altaffilmark{9} }

\altaffiltext{1}{Cluster for Pioneering Research, RIKEN, Saitama 351-0198, Japan}
\email{hirotaka.ito@riken.jp}
\altaffiltext{2}{Interdisciplinary Theoretical \& Mathematical Science Program (iTHEMS), RIKEN, Saitama 351-0198, Japan}
\altaffiltext{3}{Keio Institute of Pure and Applied Sciences, Keio University, Yokohama 223-8522, Japan}
\altaffiltext{4}{Astrophysical Big Bang Laboratory (ABBL), RIKEN, Saitama 351-0198, Japan}
\altaffiltext{5}{Astrophysical Big Bang Group (ABBG), Okinawa Institute of Science and Technology Graduate University (OIST), 1919-1 Tancha, Onna-son, Kunigami-gun, Okinawa 904-0495, Japan}
\altaffiltext{6}{Florida Institute of Technology, 150 W. University Blvd, Melbourne, FL 32901, USA}
\altaffiltext{7}{Institute of Astronomy, Russian Academy of Sciences, Moscow, 119017 Russia}
\altaffiltext{8}{Space Research Institute, Russian Academy of Sciences, Moscow, 117997 Russia}
\altaffiltext{9}{College of Science and Engineering, School of Mathematics and Physics, Kanazawa University, Kakuma, Kanazawa, Ishikawa 920-1192, Japan}

\begin{abstract}
We explore the properties of photospheric emission in the context of long gamma-ray bursts (LGRBs) using three numerical models that combine relativistic hydrodynamical simulations and Monte Carlo radiation transfer calculations in three dimensions. 
Our simulations confirm  that the photospheric emission gives rise to correlations between the spectral peak energy and luminosity that agree with the observed Yonetoku, Amati, and Golenetskii correlations. It is also shown that the spectral peak energy and luminosity correlate with the bulk Lorentz factor, as indicated in the literature. On the other hand, synthetic spectral shapes tend to be narrower than those of the observations. The result indicates that an additional physical process  that can provide non-thermal broadening is needed to reproduce the spectral features. Furthermore, the polarization analysis finds that, while the degree of polarization is low for the emission from the jet core ($\Pi < 4~\%$), it tends to increase with the viewing angle outside the core and can be as high as $\Pi \sim 20-40~\%$ in an extreme case.
This suggests that the typical GRBs show systematically low polarization compared to softer, dimmer counterparts (X-ray-rich GRBs and X-ray flashes).
Interestingly, our simulations indicate that photospheric emission exhibits large temporal variation in the polarization position angle  ($\Delta \psi \sim 90^{\circ}$), which may be compatible with those inferred in observations. A notable energy dependence of the polarization property  is another characteristic feature found in the current study. Particularly, the difference in the position angle among  different energy bands can be as large as $\sim 90^{\circ}$.

\end{abstract}


\section{INTRODUCTION}

Despite the decades of research since their discovery, the origin of the prompt emission of gamma-ray bursts (GRBs) has been a puzzle. One of the main questions  is whether the radiation is optically thin synchrotron or photospheric. Each scenario has advantages and disadvantages; therefore, the issue remains controversial. It may be possible that both emissions play a key role in the prompt phase.

Spectral features have often been invoked as important diagnostics for this issue. While synchrotron emission is favored due to
the non-thermal nature of prompt emission, it has been suggested in the literature that it fails to reproduce the low energy part and the width of spectra in many GRBs \citep[e.g.,][]{PBM98, AB15, Yu19, Acuner20, Begue20, Li21}. Combined with the fact that it is a natural outcome of the original GRB fireball model \citep{G86, P86}, these spectral analyses reactivated the study of photospheric emission as a plausible alternative. Various authors have  demonstrated that photospheric emission is capable of producing observed non-thermal features by considering dissipation \citep{PMR05, IMT07, G08, LB10, B10, VBP11, CL15, ALN15, VB16, BK20} and/or geometrical effects \citep{INO13, INM14, LPR13, LPR14} in the sub-photospheric region.

However, recently it has been claimed that  earlier critiques of synchrotron emission may be unjustified since the analysis is biased by the phenomenological fitting formula employed in the studies \citep{ONG17, ONG19, B19, BBG20}. Although there are a few exceptions, such as GRB090902B \citep{Ryde10} and 220426A \citep{Song22}, these studies argue that spectra produced by the physics-based calculation of synchrotron emission can fit the great majority of the observed GRB. It should be noted, however, that fitting requires fine-tuning of physical conditions (e.g., magnetic fields, minimum energy of electrons).\footnote{See also the discussion (Section 4.1) given in \citet{Begue20}.} It is unclear whether such conditions can be achieved in reality.

Apart from the issue of the spectral features, polarization properties can provide an independent clue to the prompt emission mechanism \citep[see, e.g.,][for a comprehensive overview]{Gill21}. Several observational efforts have been devoted to measuring the polarization signature. Early attempts by RHESSI and INTEGRAL have inferred highly polarized emission in few GRBs, but these observations seem to suffer from instrumental systematics \citep[][]{CB03, RF04, WHA04, KBK07, MCD07}. The first dedicated GRB polarimeter, the GAP instrument on board IKAROS, achieved the first polarization measurement with low systematic uncertainty \citep{YMG11a}. A high degree of linear polarization ($\sim 30-80\%$) in 3 GRBs was reported in their observations \citep{YMG11b, YMG12}.
Recent updates were provided by the second dedicated GRB polarimeter POLAR on board Tiangong-2 \citep{ZKB19, Kole20} and also by CZTI on board AstroSAT \citep{Chatto22}, although not dedicated to GRB detection. Contrary to the measurements of GAP, 14 GRBs observed by POLAR are found to be mostly compatible with a low to null polarization. AstroSAT also finds that a majority of their sample (20 GRBs) is consistent with being low to null polarization, whereas a fraction of 5 GRBs is inferred to exhibit high polarization ($\gtrsim 40~\%$). Time-resolved polarization analysis performed in these observational studies has also suggested that some GRBs \citep[GRB 110826A, 170114A, 160821A:][]{YMG11b, BKM19, Sharma19} exhibit a  significant temporal variation in the polarization position angle. It is discussed in the literature \citep{BKM19} that the low polarization inferred in the majority of GRBs detected  by POLAR and AstroSAT may result from the variation of the position angle: the emission may possess intrinsically high polarization, but it is canceled out in the time-integrated measurements as a result of the summation of polarized emissions with various position angles. All polarization measurements that are currently available, however, suffer from low photon statistics and accompanying large error bars.
Thus, a clear picture of the polarization properties is yet to be established. Future polarization missions such as POLAR-2 \citep{Kole19}, LEAP \citep{McConnel21},  and COSI \citep{Tomsick19} are awaited for further constraints.

From a theoretical perspective, many studies have explored polarization properties in the framework of optically thin synchrotron  models \citep[e.g.,][]{LPB03, G03,   NPW03, W03,  TSZ09, L06, ZY11, GGK20, GG21}. These studies have shown that, depending on the configuration of magnetic fields, the geometry of outflows, and viewing angles, various degree of polarization  ranging from $\sim 0\%$  up to $\sim 70\%$ can be produced. Several studies also  explored the  polarization in photospheric emissions. The initial study by \citet{B11} finds that, while the photons released at a local emitting region are strongly polarized, the superposition of each emission component suppresses the net polarization if the outflow is uniform. Following studies  have shown up to $\sim 30-40\%$ polarization degree can be achieved for an outflow with a sharp velocity gradient in a lateral direction \citep{INM14,  LPR14}. The study by \citet{LVB18} has also shown that if synchrotron emission is induced by sub-photospheric dissipation, the resulting polarization of the photospheric emission can be significant at energies below the spectral peak even when the outflow is uniform.

Combined with  spectral features, these predictions of the polarization properties can be used to discriminate between the emission models. It is noted, however, that most studies invoke simplified outflow structures for the theoretical modeling of the GRB emissions. To further test the emission models, it is highly desirable to conduct a study based on realistic jet dynamics. In this respect, an approach to evaluate the photospheric emission based on relativistic hydrodynamical simulation has developed in the last decade. In the early years, a series of studies have explored the properties of long GRBs (LGRBs) by applying a crudely approximated photosphere model to hydrodynamical simulations of a relativistic jet breaking out of massive stellar envelope \citep{LMB09,  LMB13, MNA11, NIK11, LML14}. Later, a radiation transfer calculation was incorporated to improve the accuracy in evaluating the emission signatures. The initial study by \citet{IMN15} applied their Monte-Carlo radiation transfer code \citep{INO13, INM14} to compute the light curves and spectra of LGRB based on hydrodynamical simulations. Following the study, several works have carried out numerical simulations of photospheric emission using a similar numerical method \citep[i.e., post-process Monte-Carlo radiation transfer calculation based on a hydrodynamical simulation of LGRB jets;][]{L16, PL18, PLL18, PL21, IMN19}. The latest simulations have updated the method to incorporate the calculation of polarization \citep{PLL20, Parsotan22}. Recently, the numerical technique was also  applied to explore the photospheric signals in short GRBs (SGRBs) \citep{IJT21}.

Although possible effects from non-hydrodynamical dissipation processes (e.g., magnetic reconnection and hadronic collisions) are not taken into account, these simulation-based studies have provided an important insight into the imprint of the realistic jet dynamics on the photospheric emission. A notable finding of these studies
is that correlations among the spectral peak energy, $E_p$, isotropic energy, $E_{iso}$, and peak luminosity, $L_p$, that are compatible with the observed GRB prompt correlations arise as a natural outcome of the simulation. While tension from the observation remains in the simulated spectral shapes, the ability to reproduce various empirical prompt correlations is regarded as one of the major advantages of the photospheric emission model \citep[see][for a recent review on the prompt correlations and their theoretical interpretations]{ParsotanIto22}.

The aim of the current paper is to gain further insight into the properties of photospheric emission in LGRBs. To this end, we extend our previous study \citep[][hereafter, IMN19]{IMN19} 
and carry out a more comprehensive analysis of the emission properties. In IMN19, we explored light curves and time-integrated spectra for three sets of simulations and mainly focused on the correlation between the spectral peak energy  $E_p$ and peak luminosity $L_p$ (Yonetoku relation). In this paper, we  further explore various prompt correlations and conduct a detailed analysis of the spectrum and polarization, which were not investigated in IMN19. While the same set of hydrodynamical simulations is considered as in IMN19, we enlarged the number of photons tracked in the Monte-Carlo radiation transfer calculation by a factor of $\sim 10$. The increase in the photon statistics is particularly important for resolving spectral shapes and polarization properties with reasonable accuracy.

This paper is organized as follows. In Section \ref{sec:model}, we describe our model and numerical procedures employed in our calculations.  In Section \ref{sec:Jetstruc}, we show the transverse structure of the jet obtained from our hydrodynamical simulations, which plays an essential role in the emission properties. Section \ref{sec:correlations} summarizes the correlations found in the time-integrated emissions. The results of the time-resolved spectral  and polarization analysis are presented in Section \ref{sec:result1} and Section \ref{sec:result2}, respectively. Section \ref{sec:summary} is devoted to a summary and discussions.

\section{MODEL AND METHODS}
\label{sec:model}

As described in IMN19, we analyze the photospheric emissions evaluated in 3 simulations that employ different jet power. The procedure is as follows: First, we compute the  evolution of a relativistic jet by performing a 3D hydrodynamical simulation. Then radiation transfer within the jet is solved as a post-process based on the Monte-Carlo method. The details of the numerical method and setup are described in IMN19. The differences from the previous paper are that we have increased the number of photon packets by an order of magnitude in each simulation (from $\sim 5\times 10^8$ to $\sim 5\times 10^{9}$) and also incorporated the polarization calculation in the radiation transfer calculation. As for the polarization calculation, we follow the  method described in \citet{IJT21} \citep[see also][for further detail]{INM14}. Here we give a brief overview of the numerical setup.

In the hydrodynamical simulations, we inject a relativistic jet inside a massive star which eventually breaks out from the stellar envelope.
The subsequent evolution is solved up to the point where it becomes optically thin to radiation.\footnote{To ensure that the jet has become optically thin even at high latitudes ($\theta_{\rm obs} \gtrsim 4^{\circ}$), we follow the evolution of our jets up to $r \sim 2\times 10^{14}~{\rm cm}$.} We consider three models that employ different jet kinetic powers: $L_{\rm j} = 10^{49}$, $10^{50}$, and $10^{51}~{\rm erg~s}^{-1}$. In all simulations, the jet is continuously injected for a duration of $t=100~{\rm s}$ with an initial opening angle of $\theta_{\rm ini} = 5^{\circ}$ and an initial Lorentz factor of $\Gamma_{\rm ini} = 5$, at the inner boundary of the simulation, which is located at a radius of $r=10^{10}~{\rm cm}$. Note that the jet expands to an opening angle of $\sim \theta_{\rm ini} + 0.7\Gamma_{\rm ini}^{-1}$ before it encounters the first recollimation shock \citep{HGN18}. The initial specific enthalpy is set as $h_{\rm ini} = 100$ for the models with $L_{\rm j} = 10^{49}$ and $10^{50} ~{\rm erg/s}$, while $h_{\rm ini} = 180$ is adopted in the model with $L_{\rm j} = 10^{51} {\rm erg/s}$. As for the progenitor star model, we employ model 16TI provided by \citet{WH06}.
The breakout of the jet head from the stellar envelope takes place approximately at $t \approx 25$, $8$, and $4~{\rm s}$ in the models with jet powers of $L_{\rm j} = 10^{49}$, $10^{50}$, and $10^{51}{\rm erg~s}^{-1}$, respectively.
In line with previous studies on collapsar jets \citep[e.g.,][]{LMB09, GLN19}, our models also demonstrate that the jet propagates at subrelativistic speeds before the breakout. 
 Nonetheless, in contrast to simulations of similar setups (jet power and stellar radius), our models exhibit a moderately reduced breakout time, which is likely due to the large jet injection radius imposed in the current study.

The Monte-Carlo radiation transfer simulation solves the evolution of the photon packets initially injected at highly optically thick regions. Each packet is treated as an ensemble of photons with the same frequency $\nu$ and, as a whole, carry the Stokes parameters: $I$, the intensity, and  $Q$ and $U$, the two parameters which characterize the linear polarization. Here, the intensity parameter is defined as $I=n_{\rm pack} h\nu$, where  $n_{\rm pack}$ is the number of photons in the packet. In the current simulation, $n_{\rm pack}$ does not vary throughout the evolution, and an equal number is imposed among all photon packets. Hence, $I$ depends only on $\nu$. From the Stokes parameters, the linear polarization can be quantified: $\Pi = \sqrt{Q^2 + U^2}/I$ and $\psi = 1/2 ~{\rm arctan}(U/Q)$ give the degree and the position angle of linear polarization, respectively. 
We do not take into account the Stokes parameter $V$, the circular polarization parameter,  because it is irrelevant in the current simulation.\footnote{As long as $V=0$ is imposed at the initial injection and the spin of the electrons have an isotropic distribution, circular polarization remains zero throughout the simulation \citep[e.g.,][]{Depaola2003}}
 Regarding the coordinate system employed to define $Q$ and $U$, we use the same prescription as in \citet{INM14} (see their Figure~4). Hence, $Q>0$ ($Q<0$) and $U=0$, i.e., $\psi = 0^{\circ}$ ($90^{\circ}$), corresponds to the case when the electric vector of the polarized beam is parallel (perpendicular) to the plane formed by the line of sight (LOS) of the observer and the jet axis. At the injection,  photon packets are set to be unpolarized ($Q=U=0$), and their frequency  is drawn randomly with a probability proportional to  the Planck distribution of the local temperature. After the injection, we self-consistently calculate the evolution of $\nu$ (or equivalently $I$), $Q$ and $U$ due to  scatterings by thermal electrons
until the packet reaches the outer boundary of the simulation.

Throughout the paper, the observer's location is expressed by the viewing angle $\theta_{\rm obs}$, which is the angle between  the LOS  and the central axis of the jet. Since our calculation is performed in 3D, the emission depends not only on the viewing angle but also on the azimuthal angle. However, the dependence turns out to be weak in the current models. Therefore,  we  show the results for a fixed azimuthal angle, i.e., all LOSs are aligned on a single plane.
In our current simulations, we adopt a spherical coordinate system ($r$, $\Theta$, $\Phi$), with the central axis of the jet aligned along the direction of $\Theta = 90^\circ$ and $\Phi = 90^\circ$. Hence, the plane of the LOS corresponds to the $\Phi = 90^\circ$ plane. Hereafter, we focus on this plane and express the zenith angle of the simulation as $\theta$, with $\theta = 0^{\circ}$ aligned to the jet axis, i.e., $\theta = \Theta - 90^{\circ}$. The viewing angle $\theta_{obs}$ is defined in exactly the same way as $\theta$, so that $\theta_{obs} = \Theta - 90^{\circ}$.

\section{Transverse structure of jet}
\label{sec:Jetstruc}

\begin{figure}[htbp]
\begin{center}
\includegraphics[width=8.5cm,keepaspectratio]{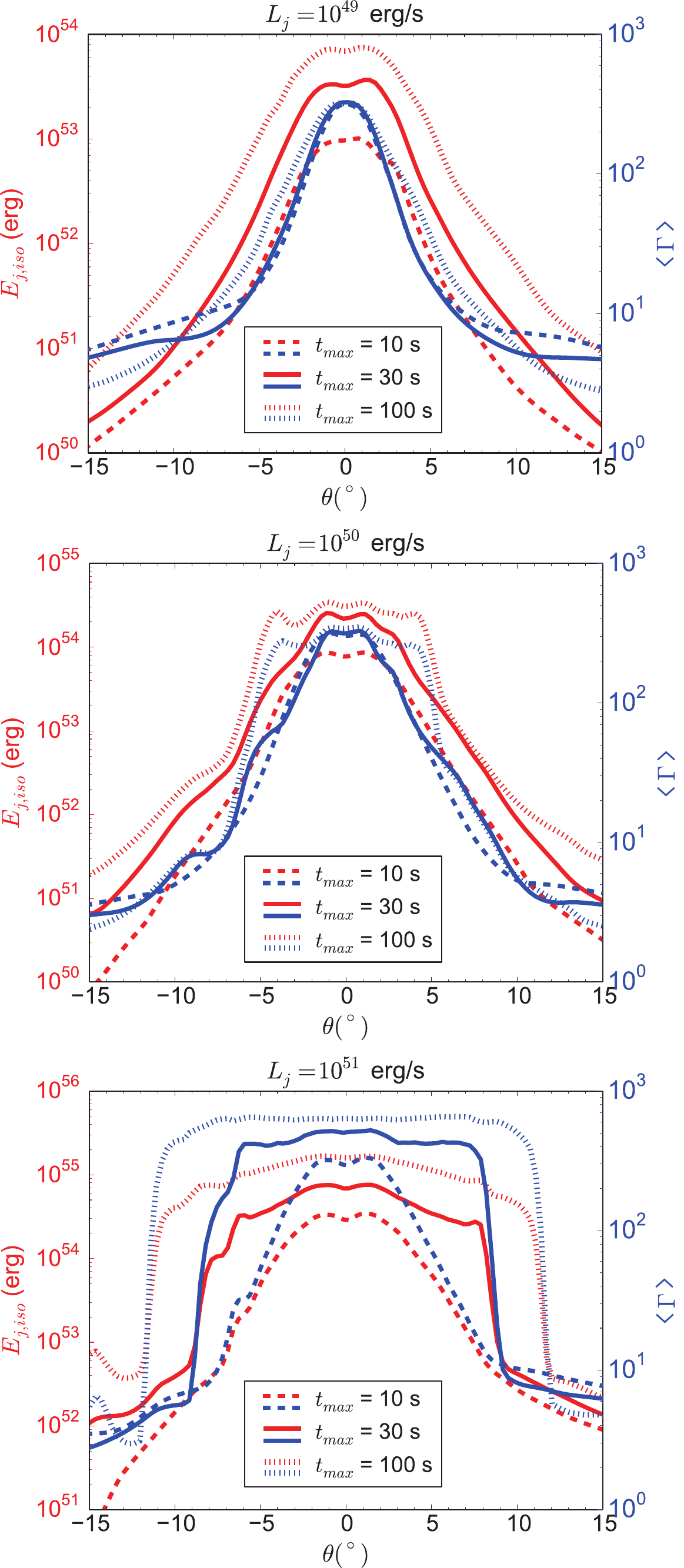}
\end{center}
\caption{Polar distribution of the accumulated isotropic energy $E_{j,iso}$ ({\it red})  and the average Lorentz factor $\left<\Gamma \right> $ ({\it blue}) of the outflow evaluated at $r=10^{13}~{\rm cm}$ in a 2D slice taken through the midplane of the hydrodynamical simulation. The dashed, solid, and dotted lines for $E_{j,iso}$ correspond to the values obtained by integrating over the duration $t_{\rm max} = 10$, $30$ and $100~{\rm s}$, respectively. The dashed, solid, and dotted lines for the Lorentz factor $\left< \Gamma \right>$ are the average  values weighted with the energy of the outflow over the same duration.}
\label{fig:1}
\end{figure}

Before presenting the result of the spectral and polarization analysis, let us give a brief overview of the transverse structure of the jet found in the hydrodynamical simulation since it plays a key role in determining the properties of the emission. In Figure \ref{fig:1}, we show the polar distribution of the accumulated isotropic equivalent energy $ E_{j,iso} $ ({\it red})  and the energy-weighted average value of the Lorentz factor $\left< \Gamma \right> $ ({\it blue}) evaluated at  $r=10^{13}~{\rm cm}$. These quantities are computed as functions of $\theta$:
\begin{eqnarray}
E_{j,iso}(\theta) = \int_{t_{13}}^{t_{13} + t_{max}} L_{j,iso}(\theta) ~dt,\\
\left< \Gamma  (\theta)\right>=  \frac{1}{E_{j,iso}(\theta)}\int_{t_{13}}^{t_{13} + t_{max}} \Gamma(\theta)~ L_{j,iso}(\theta) ~dt,
\label{Eq2}
\end{eqnarray}
where $L_{j,iso} = 4\pi r^2 \Gamma^2 [\rho c^2 (h - 1/\Gamma)] \beta_j {\rm cos}\theta_{\beta} c$ is the isotropic equivalent jet kinetic power, with $c$ being the speed of light, and $\rho$ and $\beta_j$ denoting the mass density and the velocity normalized by the speed of light, respectively.
In the above equations, all physical quantities are evaluated at a given zenith angle $\theta$ on a local mesh located at $r=10^{13}~{\rm cm}$.
Here, $\theta_{\beta}$ is the angle between the direction of velocity and radial direction. The time when the forward shock reaches $r=10^{13}~{\rm cm}$ is denoted by $t_{13}$, and the integration is performed from that time for a duration of $t_{max}$. Hence, the time integration implies the accumulation of the outflow component that passes through the surface at $r=10^{13}{\rm cm}$ for a duration of $t_{\rm max}$. This outflow component corresponds to the region emitting radiation from $t_{\rm obs}=0$ to $\approx t_{\rm max}$.
Note that the radius  $r=10^{13}~{\rm cm}$ is significantly larger than the stellar radius ($\approx 4\times 10^{10}~{\rm cm}$), implying that the jet is in the phase of free expansion after breaking out from the star.
The various lines in the figure correspond to the difference in the duration of the time integration, $t_{max}$, for the computation of  $ E_{j,iso} $ and  $\left< \Gamma \right> $. As  found in the previous studies \citep[e.g.,][]{GNB21}, the outflow structure shows a  core region at the center in which  $E_{iso, j}$ and  $\left< \Gamma \right> $ are close to uniform  that is surrounded by a wing region which shows a rapid decline in these quantities. Such a structure is formed due to the interaction between the jet and the massive stellar envelope. As shown in  IMN19 and in the current paper,   the transverse structure gives rise to the spectral-luminosity correlation.

As seen in Figure \ref{fig:1}, while the opening angle of the core, $\theta_{\rm core}$, remains nearly constant throughout the entire duration for the model with $L_{\rm j} = 10^{49}~{\rm erg/s}$, it rapidly broadens at $t \sim 30~{\rm s}$ and $\sim 10~{\rm s}$ for models with $L_{\rm j} = 10^{50}~{\rm erg/s}$ and  $10^{51}~{\rm erg/s}$, respectively. The sudden spread in these models reflects the time the innermost recollimation shock\footnote{ As found in the studies involving hydrodynamical simulations of collaspsar jets \citep[e.g.,][]{LMB09, GLN19}, the jet undergoes the formation of multiple recollimation shocks before it successfully breaks out of the stellar envelope. Here, the term "innermost" recollimation shock refers to the first shock encountered
by the jet after its injection at the base. Eventually, this innermost shock also breaks out the stellar envelope, leading to the emergence of the unshocked (uncollimated) jet from the envelope.}
breaks out from the stellar envelope; therefore, confinement by the cocoon pressure becomes less effective at later time. Before the transition, the opening angle of the core is $\theta_{\rm core} \sim 2 - 4^{\circ}$ in all models, which is  consistent with the scaling  found in previous studies, $\theta_{\rm core} \sim (\frac{1}{5} - \frac{1}{3}) (\theta_{\rm ini} + 0.7 \Gamma_{\rm ini}^{-1})$ \citep[][]{MI13, GNB21}. After the breakout of the recollimation shock, the unshocked jet matter that expands following the recollimation shock expells from the stellar envelope and can be visible in the prompt phase as shown in the previous studies  \citep[][IMN19]{IMN15} and also in the current paper.

\section{Correlations in the time-integrated emission properties}
\label{sec:correlations}
\subsection{The Yonetoku ($E_p-L_{p}$) and Amati ($E_p-E_{iso}$) correlations}
\begin{figure*}[htbp]
\begin{center}
\includegraphics[width=17.5cm,keepaspectratio]{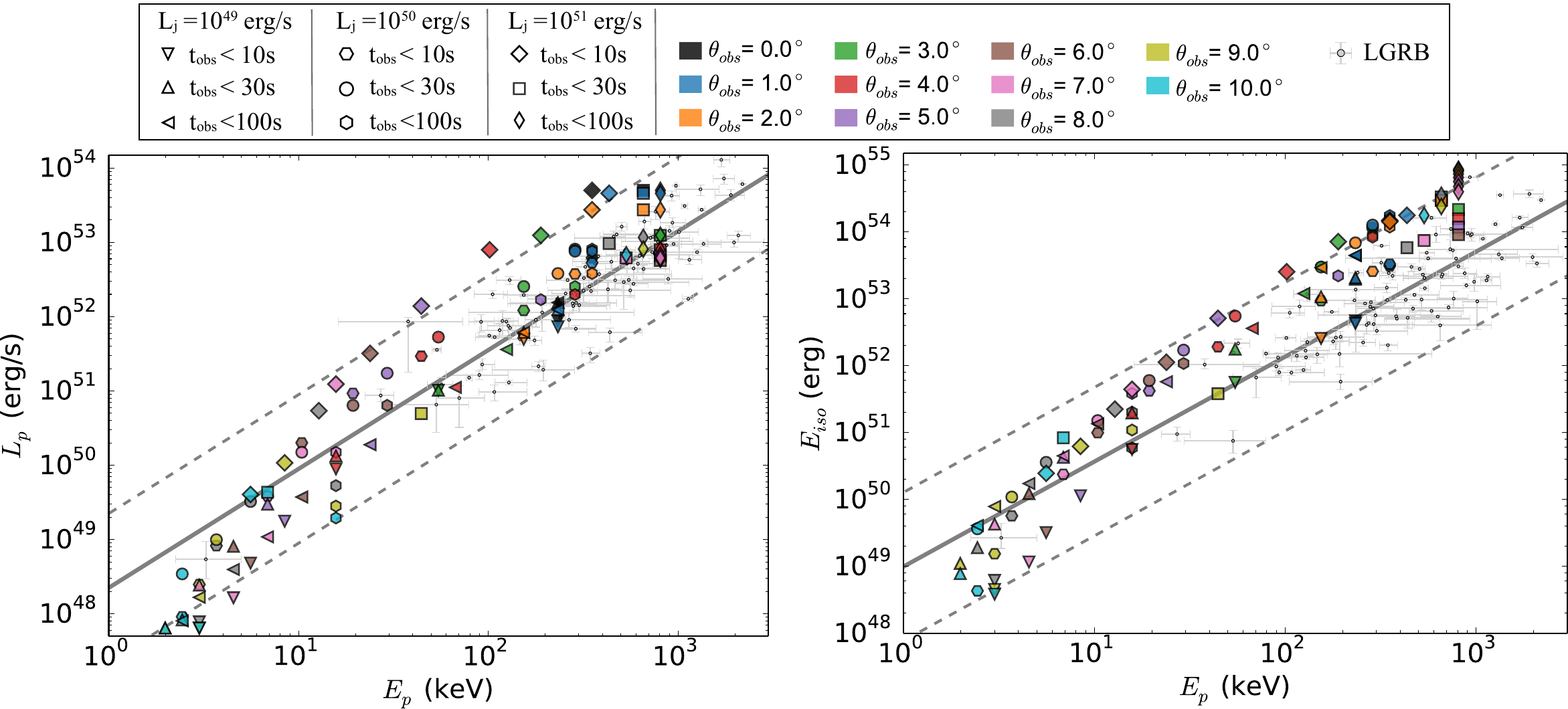}
\end{center}
\caption{Peak luminosity, $L_p$ ({\it left}), and isotropic equivalent energy, $E_{iso}$ ({\it right}), versus the peak energy, $E_p$, for different viewing angles ({\it colored symbols}).  
  The shape of the symbols represents the difference in the emission duration considered to extract $L_p$, $E_{iso}$, and $E_p$. Observational data of LGRBs taken from \citet{YMT10} (grey points) are added for comparison. The best-fitting function and $3 \sigma$ systematic error regions of the  $E_p - L_p$ and $E_p - E_{iso}$  correlations are indicated with grey solid and dashed lines, respectively.}   
\label{fig:EpLpEiso}
\end{figure*}

In IMN19, we have shown that  our simulations reproduce the correlation between the peak energy of the time-integrated spectrum, $E_p$, and the peak luminosity of the light curve, $L_p$, \citep{YMN04} quite well. We reexamine the spectral correlation with the current data sets with $\sim 10$ times better statistics. The result is displayed in the left panel of Figure \ref{fig:EpLpEiso}. As in IMN19, together with the $E_p$ and $L_p$ that are sampled from the entire emission duration (from $t_{obs} = 0$ to $t_{obs} = t_{obs, max} = 100~{\rm s}$), we also show the cases in which only emission up to a specific time ($t_{obs, max}  = 10$ and $30~{\rm s}$) is considered. This is intended to mimic bursts originating from shorter jet activity since GRBs have diverse durations. As expected, the $E_p-L_p$ correlation found in IMN19 remains unchanged,  confirming the statistical convergence of the result presented in IMN19. It is noted that the  slight change in the distribution from that displayed in IMN19 reflects the difference in the choice of the emission durations ($t_{obss,max}= 20$, $40$, and $110~{\rm s}$ are employed in Figure 3 of IMN19). In the right panel of Figure \ref{fig:EpLpEiso},  we also examine the correlation between $E_p$ and the time-integrated energy of the emission, $E_{iso}$. Our simulation results also agree with the observed $E_p-E_{iso}$ correlation \citep{Amati02}, although there is a slight offset toward higher $E_{iso}$. 

As discussed in IMN19, these correlations arise mainly due to the viewing angle effect. While the viewing angle  is within the core of the jet ($\theta_{obs} \lesssim \theta_{core}$), $E_p$, $L_p$, and $E_{iso}$ do not show significant variation. On the other hand, beyond the core ($\theta_{obs} \gtrsim \theta_{core}$), these quantities simultaneously show rapid decline with the viewing angle. As a result of the viewing dependence at the wing region, a positive correlation is found among $E_p$, $L_p$, and $E_{iso}$. Regardless of the difference in the jet power spanning two orders of magnitude, all three models agree with the observed correlations. The results are broadly consistent with the independent studies of LGRBs \citep{PL18, PLL18} and also SGRBs \citep{IJT21}, indicating that the correlations are a robust feature in the photospheric emission arising from a hydrodynamic jet. The theoretical interpretation for this outcome will be explored in future work.

\subsection{Prompt emission ($E_p$ and $L_p$) - Lorentz factor ($\Gamma_0$) correlations}

\begin{figure*}[htbp]
\begin{center}
  \includegraphics[width=17.5cm,keepaspectratio]{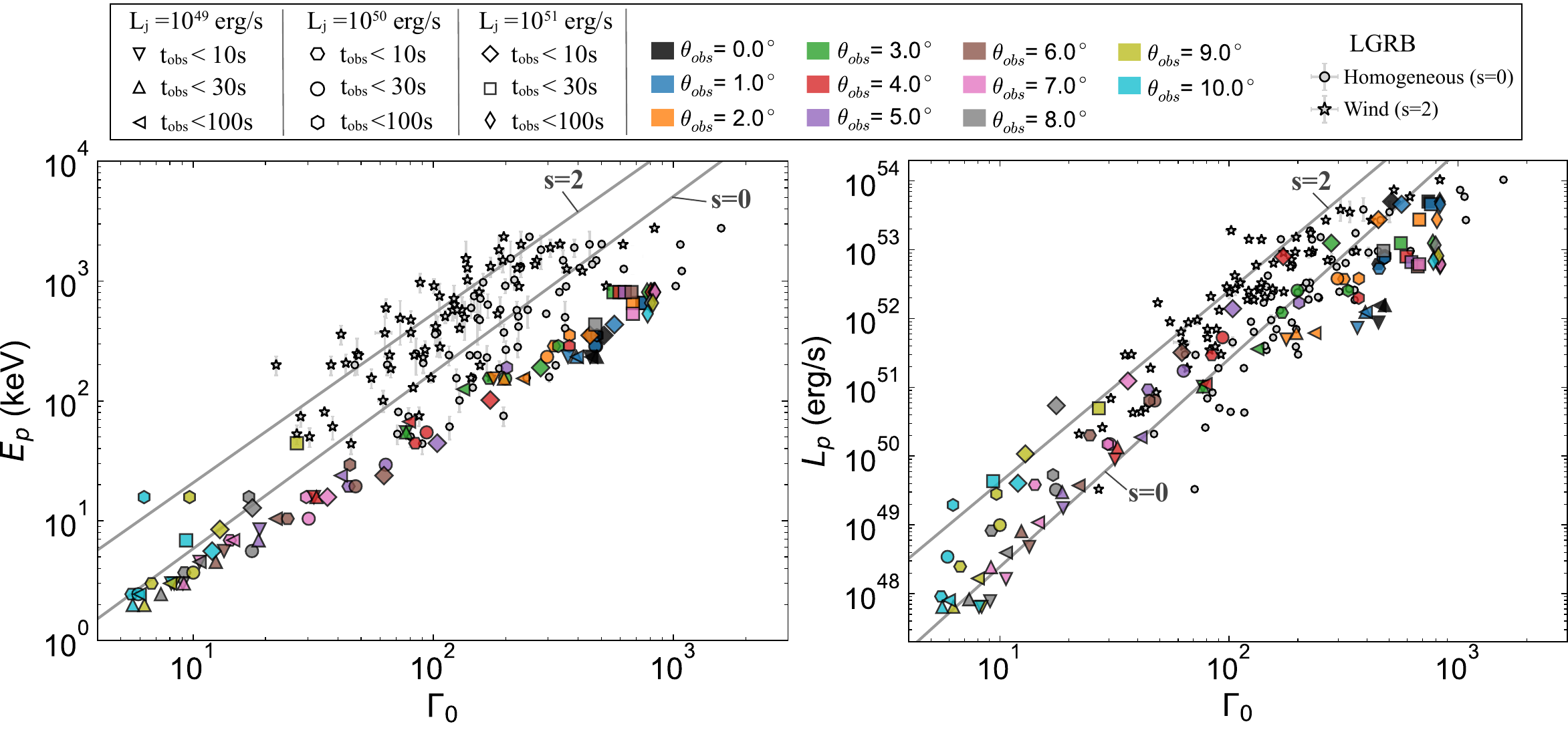}
\end{center}
\caption{Peak energy, $E_p$ ({\it left}), and peak luminosity, $L_p$ ({\it right}), versus the bulk Lorentz factor, $\Gamma_0$, for different viewing angles ({\it colored symbols}). The shape of the symbols represents the difference in the emission duration  considered to extract $E_p$, $L_{p}$, and $\Gamma_0$. 
Here, $\Gamma_0$ in our simulation is evaluated using Equation (\ref{Eq3}), and the duration of the integration, $t_{\rm max}$, is chosen to coincide with the emission duration.
For comparison, we display observed data on LGRBs from \citet{Ghirlanda18}. This study considered two different assumptions for the density profile of the circum-burst medium (CBM), which follows a form $\rho \propto r^{-s}$, to derive $\Gamma_0$. The data is depicted by grey circles for the case of homogeneous density profile (s=0) and stars for the case of a wind-like density profile (s=2). 
The best-fitting function of the $E_p - \Gamma_0$ and $L_p - \Gamma_0$ correlations for the two different assumptions are indicated with solid grey lines.}
\label{fig:EpLpGamma}
\end{figure*}

As a complementary study, we also compare our results with the empirical correlations between the emission properties ($E_p$, $L_p$, and $E_{iso}$) and the bulk Lorentz factor of the outflow, $\Gamma_0$, reported in the literature  \citep[e.g.,][]{Liang10, Liang13, Liang15, Ghirlanda12, Lu12}. Figure \ref{fig:EpLpGamma} displays $E_p - \Gamma_0$ and $L_p - \Gamma_0$ diagrams of the simulation results together with those taken from the recent observational study by  \citet{Ghirlanda18}.
In our simulations, $\Gamma_0$ is defined  as an energy-weighted average value of the terminal Lorentz factor $h\Gamma$ along the line of sight.
It is computed in the same manner as Equation (\ref{Eq2}) and can be expressed as follows: 
\begin{eqnarray}
\Gamma_0 (\theta) =  \frac{1}{E_{j,iso}(\theta)}\int_{t_{13}}^{t_{13} + t_{max}} h\Gamma(\theta)~ L_{j,iso}(\theta) ~dt.
\label{Eq3}
\end{eqnarray}
When depicting the $E_p$ ($L_p$) - $\Gamma_0$ relationship in Figure \ref{fig:EpLpGamma}, we set $\theta$ and $t_{max}$ to coincide with the viewing angle $\theta_{\rm obs}$ and the upper limit of $t_{\rm obs}$, used for determining $E_p$ ($L_p$), respectively.
As seen in the figure, our simulation shows a very good agreement with the observed $L_p - \Gamma_0$ correlation.\footnote{It is noted that, although not displayed in the figure, current simulations also show an agreement with the observed $E_{iso} - \Gamma_0$ correlation.} While there is a small offset from the center of the observed distribution, the $E_p - \Gamma_0$ diagram of our simulation also lies within the dispersion of the empirical correlation. 

It should be noted, however, that there are caveats for this comparison. One is that $\Gamma_0$ is not directly observable, and the values estimated in the literature are deduced by interpreting the hump (onset) in the early afterglow light curve as the deceleration time of the external shock. Therefore, the estimation relies on the  correctness of the afterglow model that describes the onset feature. Hence, the indirect measurement has an uncertainty associated with any possible ambiguity in the theoretical modeling \citep[see, e.g., footnote 16 in][]{ParsotanIto22}. The possible error due to ambiguity in the modeling is partially visible in Figure \ref{fig:EpLpGamma}. 
As seen in the figure,  different assumptions about the circum-burst medium (CBM), i.e., homogeneous (grey circles) or wind-like (grey stars), result in a systematic shift in the distribution.

Another caveat is on the estimation of $\Gamma_0$ based on the simulation. Here we have evaluated the quantity as an energy-weighted average value of the terminal Lorentz factor, $h \Gamma$,  along the line of sight. Regarding the core region of the jet, the value is likely to result in an overestimate (by a factor of $\sim 2$ at most for the current set of simulations) since the photosphere is located in the region at which the flow is still in the acceleration phase.\footnote{Beyond the photosphere the acceleration ceases due to the release of the radiation pressure. However, it is difficult to accurately determine the end of the acceleration phase since the location of the photosphere is fuzzy \citep{B11}.}
On the other hand, the photosphere is located above the acceleration region in the wing. Hence, one can safely determine the terminal Lorentz factor. However, since $\Gamma_0$ sharply drops with angle, it is not clear whether the onset time measured by the observer viewing the wing region ($\theta_{obs} > \theta_{core}$) reflects the deceleration time of the jet matter along the line of sight. To remove the theoretical uncertainty, the dynamics and the emission of the structured jet  during the onset of the deceleration must be solved. This is beyond the scope of this study. 

Nevertheless, keeping the above uncertainties in mind, we can conclude that no severe tension exists between the correlations found in the observations and the simulation. A similar result is obtained in \citet{LMB13}, which finds  a broad agreement between the observed $E_{iso} - \Gamma_0$ correlation and that evaluated based on the simulation. The current study, which employs a more accurate calculation method, confirms their conclusion that  photospheric emission naturally gives rise to the correlation between $\Gamma_0$ and the prompt properties.

\section{Time-resolved SPECTRAL ANALYSIS}
\label{sec:result1}

\begin{figure*}[htbp]
\begin{center}
\includegraphics[width=17cm,keepaspectratio]{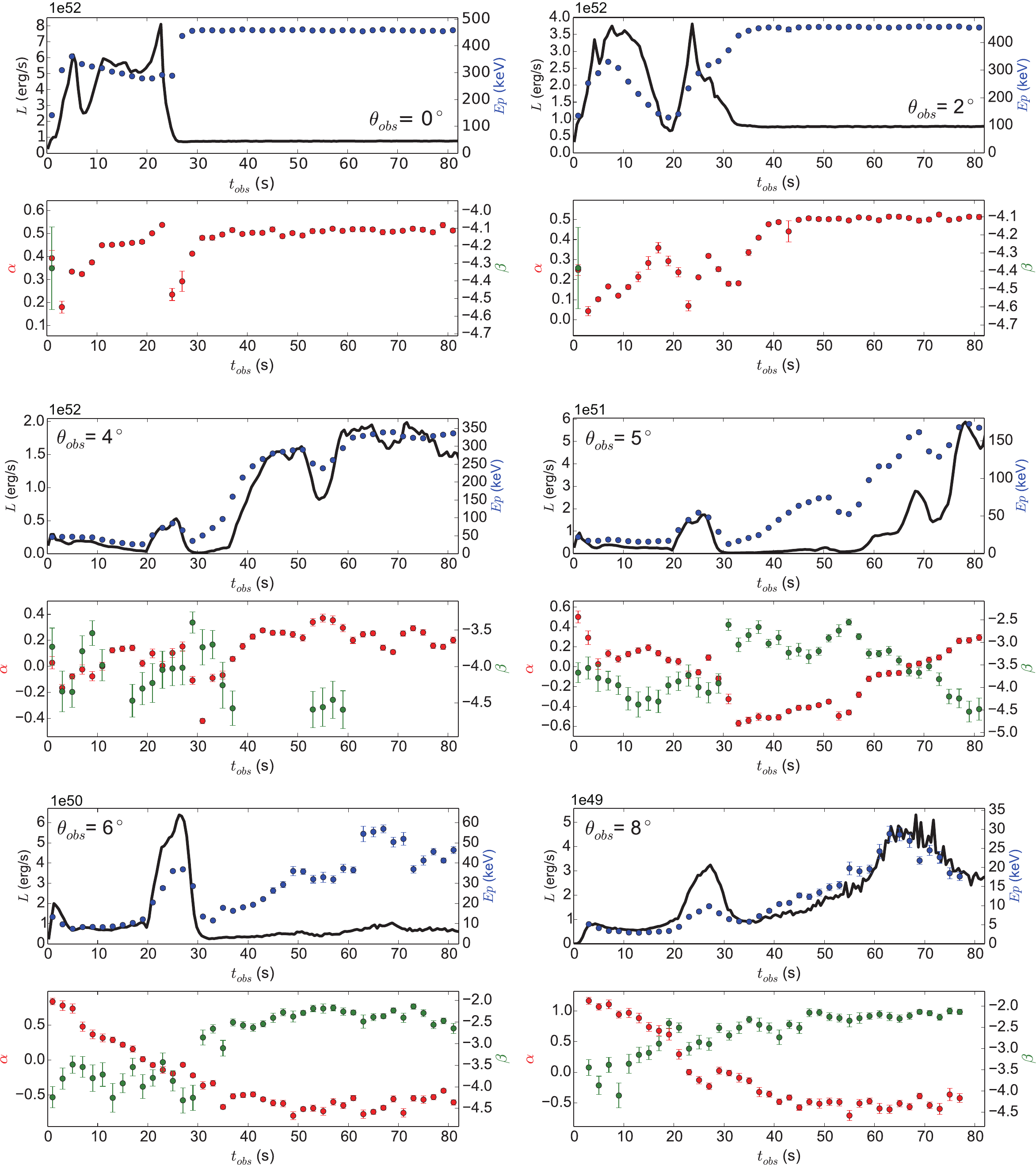}
\end{center}
\caption{Light curves and evolutions of fitted spectral parameters at various viewing angles for the fiducial model ($L_j = 10^{50}$~erg/s). The solid black lines show the light curves with time bins of $\Delta t =0.5~{\rm s}$, while the blue, red, and green circles plot the best-fit values for the spectral peak energy $E_p$, low energy photon index $\alpha$ and high energy photon index $\beta$ measured in time bins of  $\Delta t =2~{\rm s}$, respectively. Note that the parameter $\beta$ is not present in the time intervals where the spectrum is fitted using the CPL function.
}
\label{fig:Lspec}
\end{figure*}

In this section, we present the result of the time-resolved analysis. The analysis is similar to those performed by \citet{PL18, PLL18}. We employ time bins of $\Delta t = 2{\rm s}$ and all spectra are fitted either by Band function \citep{Band93} or cutoff power law (CPL) function \citep{YPG16} using the $\chi^2$ minimization method.  Here, we do not perform a goodness-of-fit evaluation between the two functions. Instead, we adopt the following procedure.
In all cases, we first fit the spectrum with Band function, which yields the peak energy, $E_p$, and the photon indices below and above the peak energy, $\alpha$, and $\beta$. When the high energy photon index is smaller than a certain value ($\beta \lesssim -4.5$), the spectral shape above the peak is mostly consistent with an exponential cut-off. For such cases, we employ the fit by the simpler CPL function, which yields $E_p$ and $\alpha$. 
In Figure \ref{fig:Lspec}, we display the results of the spectral fits for the fiducial model ($L_{\rm j} = 10^{50}~{\rm erg/s}$).

\subsection{Correlation between  $E_p$ and $L$}
\label{sec:Ep-L}
As shown in the previous studies, the peak energy $E_p$ and luminosity $L$ tend to decrease with the viewing angle due to the jet structure formed by the interaction with the stellar envelope. The emission at $\theta_{\rm obs} \lesssim 3^{\circ}$ does not show a significant difference in $E_{p}$ and $L$ since it originates from the core region of the jet in which the overall jet kinetic power and Lorentz factor are comparable. On the other hand,  a sharp decline of these quantities is seen at a larger viewing angle, reflecting the drop in kinetic power and Lorentz factor in the wing region (see middle panel of Figure \ref{fig:1}).

As seen in Figure \ref{fig:Lspec}, time evolutions of peak energy and luminosity generally show a correlation ($E_p -L$ tracking behavior). The clear exception is found at $t_{\rm obs} \sim 20-30~{\rm s}$ for $\theta_{\rm obs} \lesssim 3^{\circ}$, where the rapid drop in luminosity is accompanied by the hardening of $E_{p}$. Thereafter,  variability in the emission also vanishes. This sudden transition in the behavior is due to the fact that, while earlier emission originates from a front portion of the jet which is subject to shocks and mixing due to interaction with a dense stellar envelope, the later emission comes from the inner unshocked jet which is causally disconnected to the exterior gas \citep[for detail, see][IMN19]{IMN15}. For a wider viewing angle ($\theta_{\rm obs} \gtrsim 3^{\circ}$), the sharp transition is not present since all the emissions are from the shocked region. Nonetheless, the later emission tends to be harder because the jet-stellar interaction is weaker in the later phase.

The other two models also show similar behavior. However, the transition is less prominent even at the core region ($\theta_{\rm obs} \lesssim 3^{\circ}$) in the lower power jet model ($L_{\rm j} = 10^{49}~{\rm erg/s}$). This is because  the entire part of the jet is subject to shocks and mixing. On the other hand, the sharp transition is present for higher jet power ($L_{\rm j} = 10^{51}~{\rm erg/s}$), but the transition occurs at an earlier time ($t_{\rm obs} \sim 10~{\rm s}$). Also, since there is a rapid broadening of the unshocked jet after the transition in the high-power jet model (see bottom panel of Figure \ref{fig:1}), a steady hard emission can be observed up to  $\theta_{\rm obs}\sim 10^{\circ}$.

It is noted, however,  that such a transition may not be present in the observations \citep[see][for an attempt to identify the transition feature in the GRB light curves]{YYS17}. According to  the recent simulation by \citet{GLN19}, which solves the jet dynamics from  deeper inside the star with a higher spatial resolution, the recollimation shock does not breakout from the stellar envelope within a reasonable range of parameters. Their result suggests that the time of the transition found in the current simulation underestimates the actual value due to the large injection radius and lack of spacial resolution. Hence, the transition is unlikely to occur within the intrinsic durations in  the majority of GRBs \citep[e.g.,][]{ZGA18}. This suggests that the global hardening should be insignificant; therefore, the $E_p-L$ tracking feature is expected to be more robust throughout the emission.

\begin{figure}[htbp]
\begin{center}
\includegraphics[width=8.5cm,keepaspectratio]{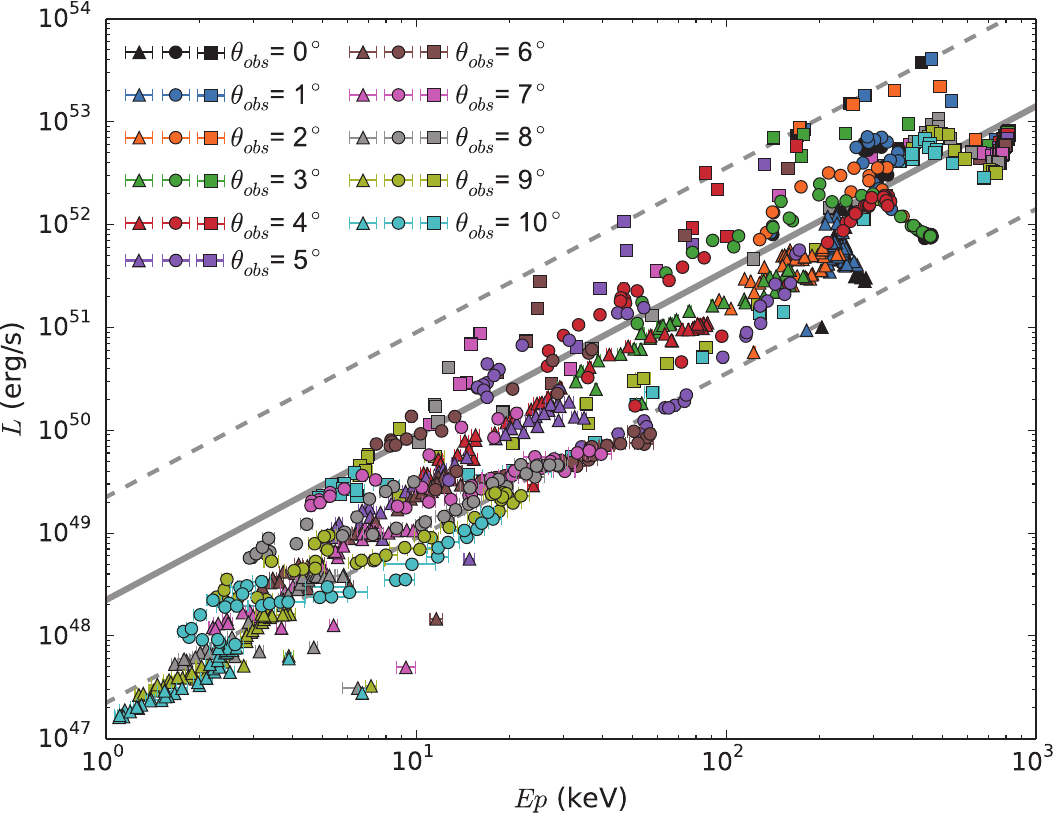}
\end{center}
\caption{The relation between spectral peak energy $E_p$ and luminosity $L$ measured in time bins of $\Delta t = 2~{\rm s}$ for various viewing angles. The triangle, circle, and square plot the results for the simulation with jet power of $L_j = 10^{49}$, $10^{50}$ and $10^{51}~{\rm erg/s}$, respectively. 
}
\label{fig:E-Lcor}
\end{figure}

As found in previous studies \citep{PLL18}, the correlation between $E_p$ and $L$ shows a good agreement with the observations \citep{GMA83, Ghirlanda10, LWL12}.\footnote{On the other hand, we found no clear evidence of the hard-to-soft pattern inferred in literature \citep[e.g.,][]{LWL12}. This result is also consistent with the independent study by \citet{PLL18}.} In Figure \ref{fig:E-Lcor}, we plot the time-resolved $E_p$ and $L$ of the three simulations. It is indicated in \citet{Ghirlanda10, LWL12} that the relation between $E_p$ and $L$ of the time-resolved spectra is consistent with that for the time-integrated spectra among the GRBs \citep[Yonetoku relation;][]{YMN04, YMT10}. The figure shows that our simulation results are in good agreement with the relation. While the viewing angle dependence gives rise to the overall correlation, the $E_p - L$ tracking feature at a given viewing angle (single burst) also roughly follows the correlation. Note that the observational data are sparse at $L \lesssim 10^{50}~{\rm erg/s}$ where the simulation results, and possibly also observations, depart slightly from the predictions of the Yonetoku relation.
Our simulation data, when fitted to the function $L = K {E_p}^{s}$, yields a best-fit value of $s = 1.87 \pm 0.01$. 
The slope is slightly steeper than that of the Yonetoku relation, which is reported as $s = 1.6 \pm 0.082$.
When excluding the data with low $L$ or $E_p$ values, the slope aligns more closely with the Yonetoku relation. For instance, if we discard the data where $L < 10^{50}~{\rm erg/s}$, the best-fit slope adjusts to $s = 1.52 \pm 0.03$.
As discussed in IMN19, this may indicate that the slope of the correlation curve  changes at low luminosities.

\subsection{Spectral shapes}
\label{sec:spectral}

As found in the time-integrated analysis \citep[][IMN19]{IMN15}, the time-resolved spectra tend to broaden with viewing angle  (Figure \ref{fig:Lspec}). Hence, the fitted value of the low energy photon index $\alpha$ tends to decrease with viewing angle. The resulting range of $\alpha$ is softer than that for a blackbody  ($\alpha = 1$). Still, it is harder than the typical observed value 
\citep[the time-resolved analysis of][found a mean and median value of $\alpha \sim -0.77$]{YPG16}
\citep[mean and median value of $\alpha \sim - 0.77$ is found in the time-resolved analysis by][]{YPG16} 
The spectrum above the peak is mostly consistent with exponential cutoff at the core part of the jet ($\theta_{\rm obs} \lesssim 3^{\circ}$), while a hard energy extension appears at a larger viewing angle. Therefore, most of the time bins are fitted by the CPL function for the former case, while Band function is used for the latter case. 
Note that the high-energy photon index $\beta$ (indicated by the green circle) is absent from the figure for the time intervals where the spectrum is fitted using the CPL function.

\begin{figure*}[htbp]
\begin{center}
\includegraphics[width=17cm,keepaspectratio]{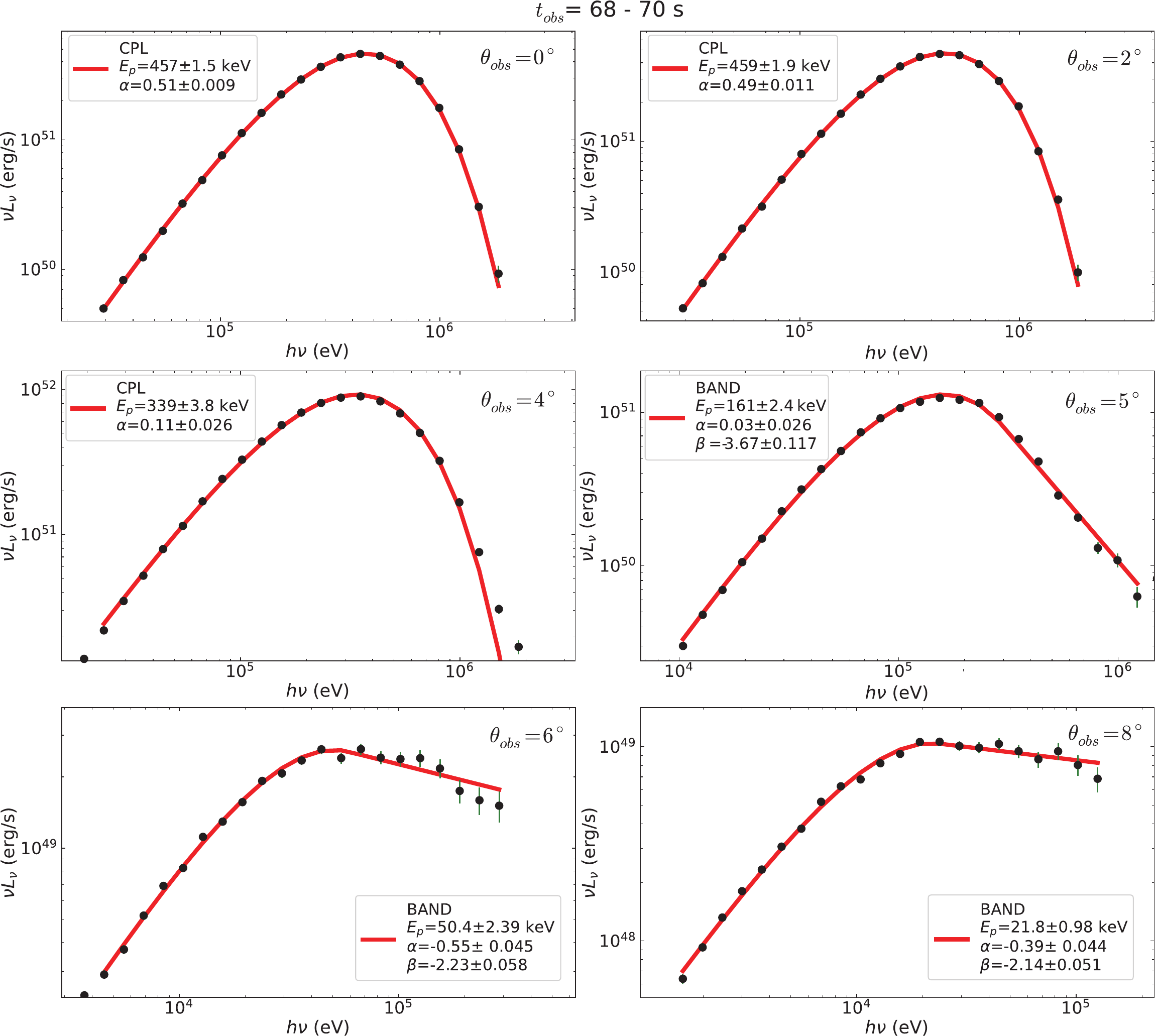}
\end{center}
\caption{Observed spectrum of the fiducial 
model ($L_j = 10^{50}~{\rm erg/s}$) at $t_{\rm obs}=68-70~{\rm s}$
for various viewing angle. The black circles represent the simulation data, and the error bars display the corresponding $1\sigma$ statistical error due to Poisson noise.}
\label{fig:SPFIT}
\end{figure*}

For illustrative purposes,  we plot the  spectrum of the fiducial model ($L_{\rm j} = 10^{50}~{\rm erg/s}$) at a specific time bin for various viewing angles in Figure \ref{fig:SPFIT} together with the corresponding best fit analytical function (either CPL or BAND). The shift toward a broader spectrum at a larger viewing angle can be confirmed in the figure.
As described in previous work \citep[][IMN19]{IMN15}, inhomogeneities in the outflow cause the spectral broadening via the multi-color effect and bulk Comptonization.  These effects combine to soften the low-energy spectrum and harden the high-energy spectrum \citep{LPR13, INO13}.


\begin{figure}
\begin{center}
\includegraphics[width=8.5cm,keepaspectratio]{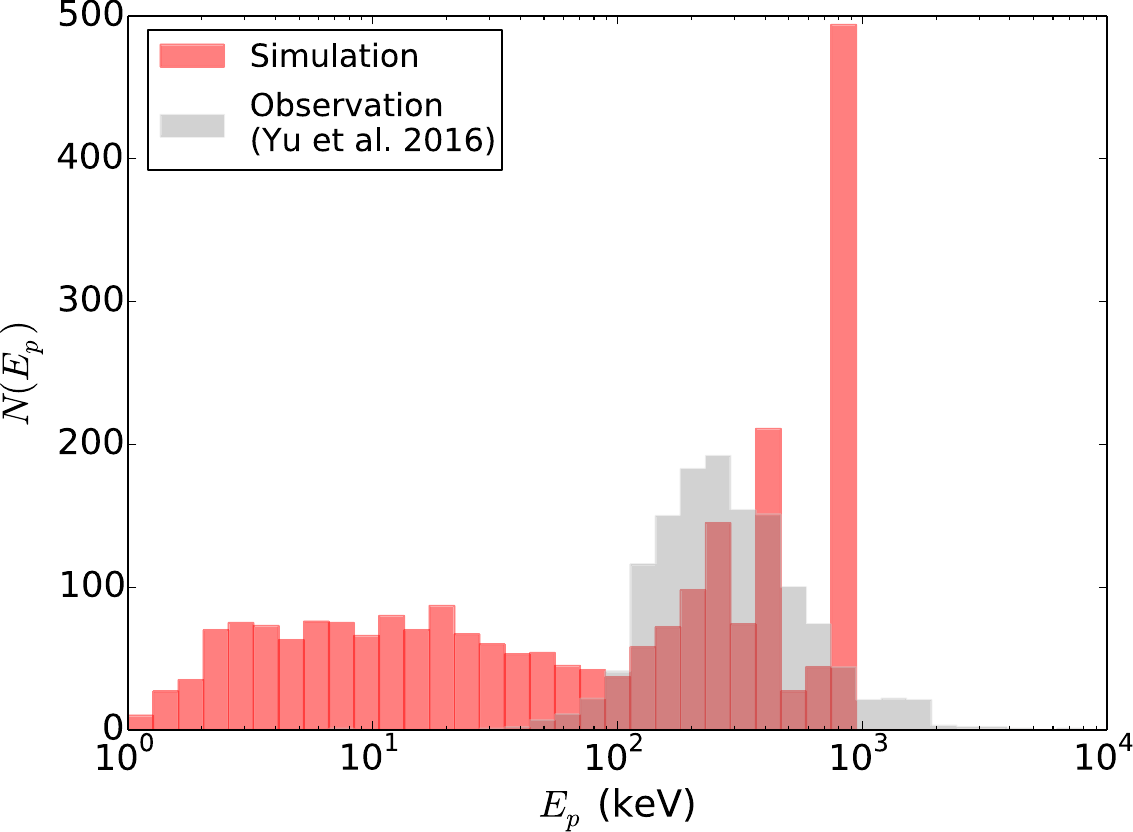}
\includegraphics[width=8.5cm,keepaspectratio]{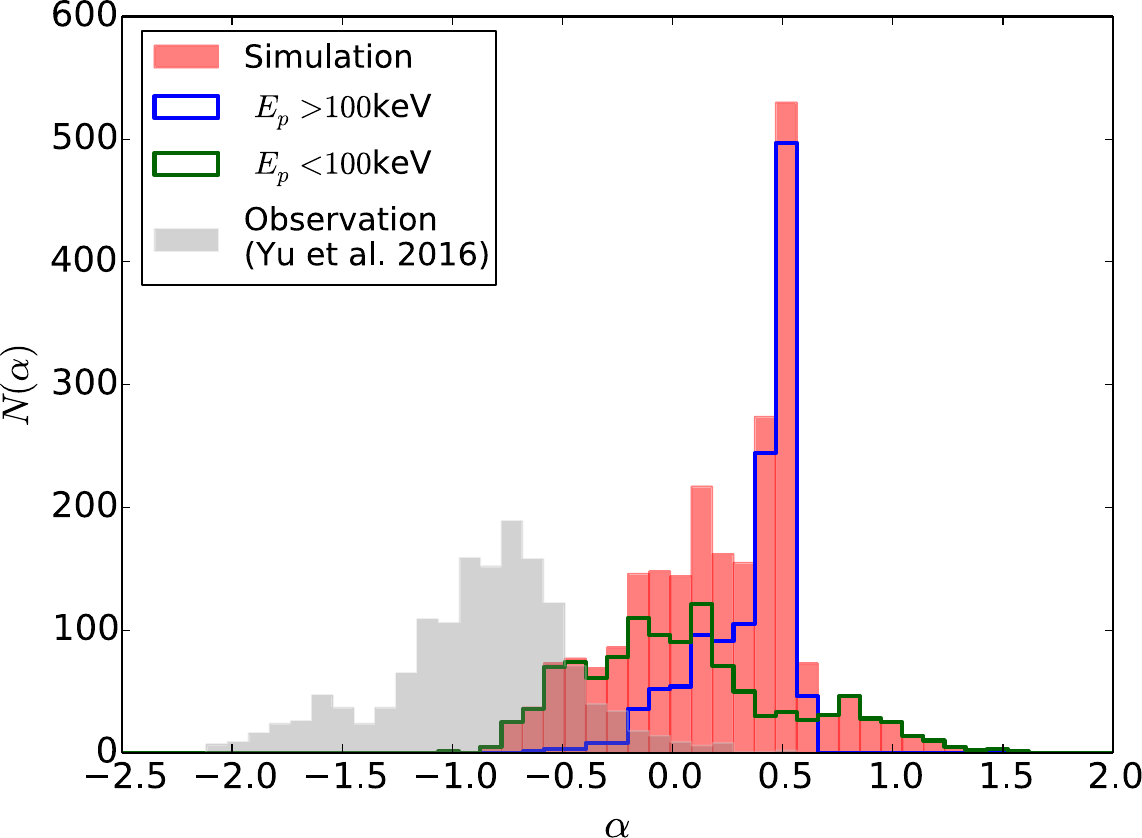}
\includegraphics[width=8.5cm,keepaspectratio]{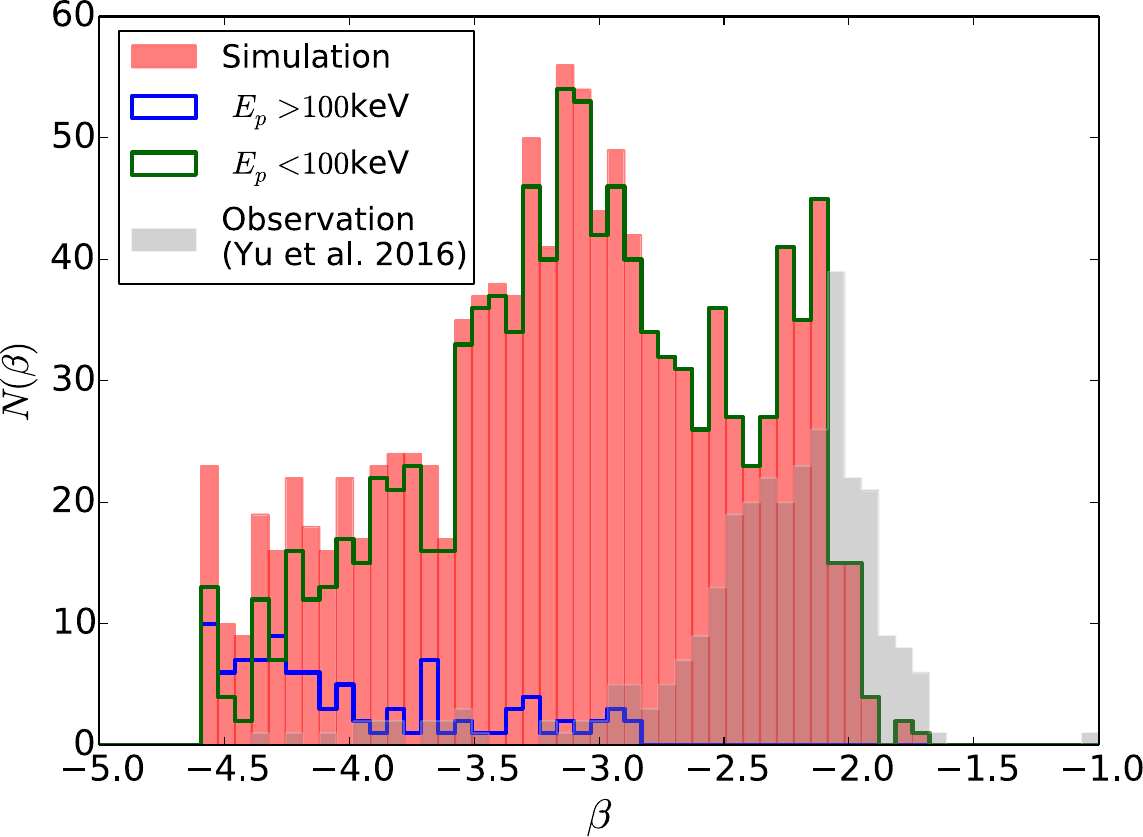}
\end{center}
\caption{Distributions of the best-fit parameters for the time-resolved spectra. 
The top, middle, and bottom panels show the spectral peak energy $E_p$, the low energy photon index $\alpha$, and the high energy photon index $\beta$, respectively. The red histograms summarize the spectra of all simulations. The considered viewing angle spans from $\theta_{\rm obs} = 0^{\circ}$ to $10^{\circ}$ with an interval of $1^{\circ}$. The grey histograms are the fitted observed spectral parameters from \citet{YPG16}. The blue and green lines in the middle and bottom panels distinguish the population of the simulated spectrum with peak energy higher and lower than $100~{\rm keV}$, respectively.}
\label{fig:FITdata}
\end{figure}

We summarize the best-fit spectral parameters for all simulation data in Figure \ref{fig:FITdata}. The broad distribution of the peak energy ranging from $\sim ~{\rm keV}$ up to $\sim 10^3~{\rm keV}$ mainly reflects the wide range of the viewing angles taken into account in the current study (see Figure \ref{fig:E-Lcor}). Note that the clustering at the highest energy $E_{\rm p} \sim 800-900~{~\rm keV}$ arises from the simulation of the highest jet power $L_{\rm j} = 10^{51}~{\rm erg /s}$. As mentioned earlier (Section \ref{sec:Ep-L}),   quasi-steady  emission is seen at a later time ($t_{obs}\gtrsim 10~{\rm s}$) in a wide range of viewing angles in this simulation. As a result, the data accumulates to produce the peak at the high-energy end of the distribution.

Examining the distribution of spectral indices, we observe that our simulations exhibit a narrow spectrum, characterized by larger values of $\alpha$ and smaller values of $\beta$, in contrast to what is typically observed.
While the observation shows clustering around $\sim -0.7$ for $\alpha$, our simulations find a sharp peak at around $0.5$ above which the number of data drops sharply. The existence of such a cutoff-like feature at $\alpha \sim 0.5$ is the expected result for non-dissipative photospheric emission from a spherical outflow \citep{B10, BSV13, LPR13, INO13}. 
This implies that the multi-color temperature effect is not significant for the population  around  the cutoff. The high energy photon index is typically in the range $-2.5 \lesssim \beta \lesssim -2.0$ for observation, while the simulation results show a broader distribution that extends down to smaller values. The broad extension in $\beta$ arises due to the gradual hardening in the high energy spectra with viewing angle (see Figure \ref{fig:SPFIT}).

While there is a clear gap between the distributions of $\alpha$ and $\beta$ for the simulation and those for the observation, some fraction shows an overlap. It is noted, however, that most of the overlapping population in the simulation corresponds to the emissions originating from a wing region that exhibits low $E_{p}$ ($< 100{\rm keV}$). Therefore, while this population may be compatible with  X-ray-rich (XRR) GRBs or X-ray flashes (XRF)  \citep{KST20}, it does not match with the typical GRB observations. 

The tension in the spectral shapes is commonly found in the simulation-based study of photospheric emission.  As discussed in our previous studies \citep[][IMN19]{IMN15}, calculation with higher spatial resolution may reduce the tension by resolving sharp shear flows, which can enhance the effects of the bulk Comptonization and  the multi-color effects. However, while the tension in the high energy photon index may be reduced by introducing intermittent central engine activity \citep{PLL18},  similar results are found in the higher resolution studies in 2D \citep[][]{L16, PL18, IJT21, PL21}. Although further 3D study is necessary to confirm this issue, these results suggest that additional physical processes not captured in the global hydrodynamical simulation must be invoked to reproduce the broad spectrum. A potential solution to this problem is a dissipative process that acts around the photosphere. As discussed in \citet{GLN19, GBL21}, the spectrum can be broadened by internal shocks that form due to rapid variability that is smeared out in the simulations \citep{BML11, ILS18, SLR22, SR22}. Moreover, additional internal dissipation that results in a significant non-thermal broadening may arise from other processes, such as magnetic reconnection and hadronic collision \citep[e.g.,][]{VB16}. The quantitative estimation of such effects  is currently infeasible and is beyond the scope of this study.

\section{Polarization Analysis}
\label{sec:result2}

In this section, we present the result of the polarization analysis. It should be  emphasized that this is the first study in which the properties of the   polarization are examined based on 3D simulations. The imposition of 2D axisymmetry which is often invoked in the literature \citep{LPR14,INM14,PLL20, IJT21, Parsotan22}, restricts the position angle $\psi = 1/2 ~{\rm arct}(U/Q)$ to be one of  two orthognal angles, i.e., $0^{\circ}$ ($180^{\circ}$)  or $90^{\circ}$ for the current definition of the Stokes parameter (Section \ref{sec:model}),  and also forces $\Pi$ at $\theta_{obs} = 0^{\circ}$ to vanish. Therefore, a study in 3D is essential to remove these artificial effects on the results of polarization analysis.

\subsection{Time-integrated polarization}
\label{sec:PDPAtint}

\begin{figure*}[htbp]
\begin{center}
\includegraphics[width=18cm,keepaspectratio]{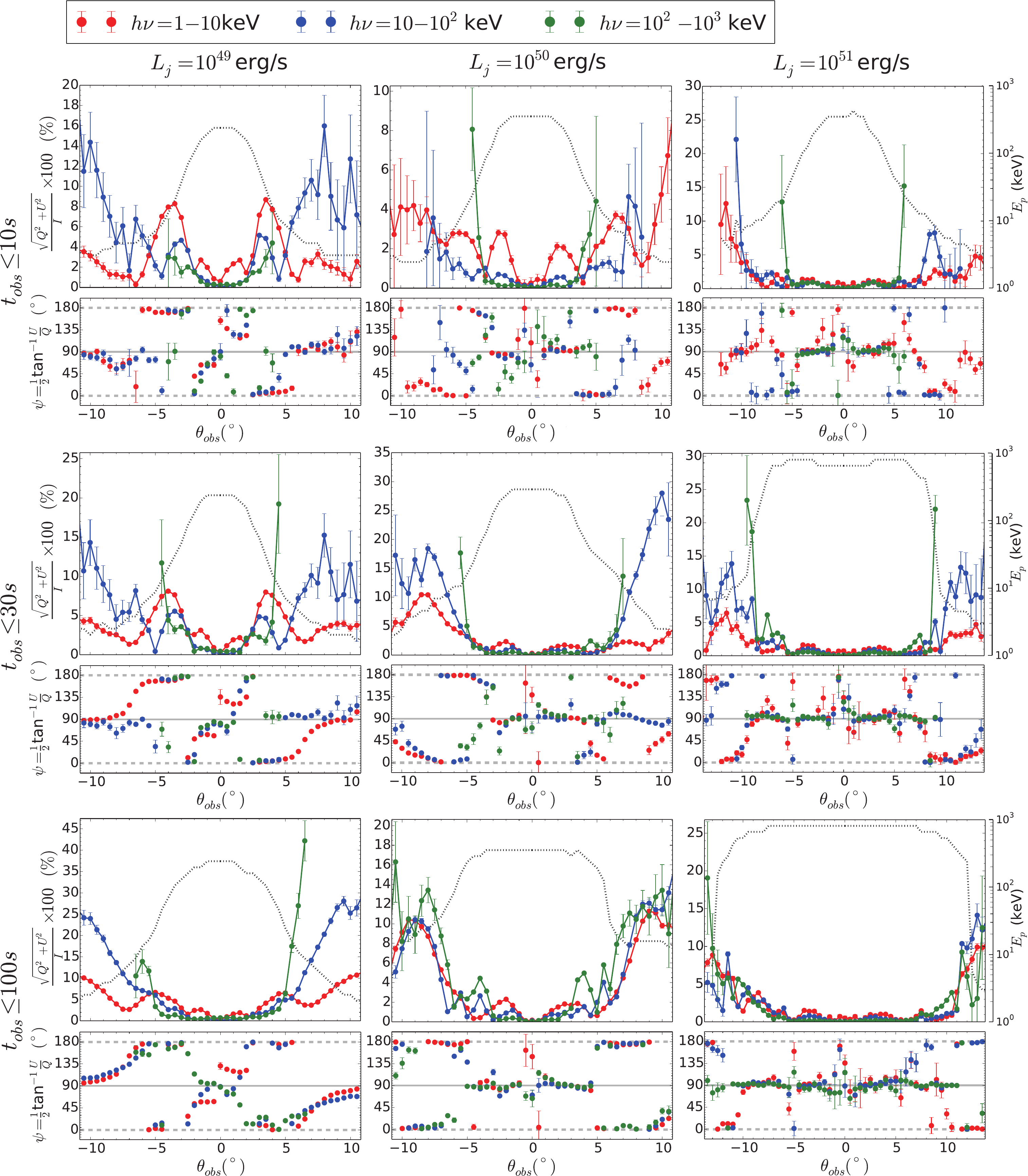}
\end{center}
\caption{
Degree of linear polarization ($\Pi = \sqrt{Q^2 + U^2}/I$ : {\it upper panels}) and position angle ($\psi = 1/2 ~{\rm arctan}(U/Q)$ : {\it lower panels}) as a function of viewing angle at energy ranges of $h\nu = 1 - 10~{\rm keV}$ ({\it red}), $10-100~{\rm keV}$ ({\it blue}), and  $100-1000~{\rm keV}$ ({\it green}). 
The error bars for $\Pi$ and $\psi$ represent the $1\sigma$ statistical uncertainty, which is determined using Equations (\ref{eq:pierr}) and (\ref{eq:psierr}). For reference,  the black dotted lines display spectral peak energy $E_p$.  The left, middle, and right column corresponds to the simulation with jet power of $L_j = 10^{49}$, $10^{50}$ and $10^{51}~{\rm erg/s}$, respectively. The top, middle, and the bottom row shows the cases in which emission up to $t_{obs} = 10$, $30$, and $100~{\rm s}$ are taken into account, respectively. We discard the viewing angles at which fewer than 100 packets were observed because of the potential for large statistical errors. Hence, the limited range of viewing angles for some curves reflects the paucity of photon packets. As for the position angle, we also discard several plots  which show $1\sigma$ error bar larger than $40^\circ$. Note that the scale of the vertical axis of $\Pi$ differs among the panels.}
\label{fig:PDPAdist}
\end{figure*}

In Figure \ref{fig:PDPAdist}, we show the time-integrated properties of polarization at three different energy bands  ($h\nu = 1 - 10$, $10-100$, and  $100-1000~{\rm keV}$) as a function of viewing angle. Following the analysis of the prompt correlations (Section \ref{sec:correlations}), we  consider three cases for the emission duration for our time-integrated polarization analysis: $t_{obs} = 0~{\rm s}$ up to $t_{obs, max} = 10$, $30$, or $100~{\rm s}$. 
In the figure, the $1\sigma$ statistical uncertainty of the degree of polarization ($\Pi = \sqrt{Q^2 + U^2}/I$) and position angle ($\psi = 1/2 ~{\rm arctan}(U/Q)$) is determined using error propagation. 
Specifically, we use the variance and covariance of the Stokes parameters $I$, $Q$, and $U$, denoted by $\sigma_{I/Q/U}^2$ and $\sigma_{IQ/IU/QU}$, to quantify the variance of $\Pi$ and $\psi$ as 
\begin{eqnarray}
\sigma_{\Pi}^2 = 
&& \left(\frac{\partial \Pi}{\partial I}\right)^2 \sigma_I^2 + \left(\frac{\partial \Pi}{\partial Q}\right)^2 \sigma_Q^2 + \left(\frac{\partial \Pi}{\partial U}\right)^2 \sigma_U^2 \nonumber \\  
&&  + ~2 \left(\frac{\partial \Pi}{\partial I}\right) \left(\frac{\partial \Pi}{\partial Q}\right) \sigma_{IQ} + 2 \left(\frac{\partial \Pi}{\partial I}\right) \left(\frac{\partial \Pi}{\partial U}\right) \sigma_{IU} \nonumber \\
&& + ~2 \left(\frac{\partial \Pi}{\partial Q}\right) \left(\frac{\partial \Pi}{\partial U}\right) \sigma_{QU} ,
\label{eq:pierr}
\end{eqnarray}
and
\begin{eqnarray}
\sigma_{\psi}^2 = 
&&\left(\frac{\partial \Pi}{\partial I}\right)^2 \sigma_I^2 + \left(\frac{\partial \Pi}{\partial Q}\right)^2 \sigma_Q^2 + \left(\frac{\partial \Pi}{\partial U}\right)^2 \sigma_U^2 \nonumber \\ 
&& + ~2 \left(\frac{\partial \Pi}{\partial I}\right) \left(\frac{\partial \Pi}{\partial Q}\right) \sigma_{IQ} + 2 \left(\frac{\partial \Pi}{\partial I}\right) \left(\frac{\partial \Pi}{\partial U}\right) \sigma_{IU}  \nonumber \\
&& +~2 \left(\frac{\partial \Pi}{\partial Q}\right) \left(\frac{\partial \Pi}{\partial U}\right) \sigma_{QU},
\label{eq:psierr}
\end{eqnarray}
respectively.
To visualize the change in the spectral properties with viewing angle, we also display $E_p$ in the figure. As described in the previous section, the viewing angle beyond which $E_p$ shows rapid decrease marks the transition from the core to the wing region as the origin of the emission.

As seen in the figure, when the observer LOS is within the core region ($\theta_{obs} \lesssim \theta_{core}$), the degree of polarization, $\Pi$, exhibits a low value: $\sim 4\%$ at most regardless of the energy band. Once the viewing angle enters the wing region ($\theta_{obs} \gtrsim \theta_{core}$), $\Pi$ tends to increase. A similar trend is found in the 2D simulation of SGRBs conducted in \citet{IJT21} \citep[see also][]{Parsotan22}. As discussed in that paper, the sharp lateral gradient of the outflow structure in the wing region results in an enhanced level of anisotropy of emission around the LOS. Hence, the polarization signature at the wing turns out to be larger than that at the core region. The main difference from \citet{IJT21} is that the current simulations have more complex viewing angle dependence (wiggling features). This presumably reflects the difference between the nature of 3D and 2D simulation, in which the former shows more complex flow structures. It is also worth noting that in the current 3D simulations, $\Pi$ exhibits a non-zero value at $\theta_{obs}=0^{\circ}$, although small ($\lesssim 1~\%$), indicating a departure from axisymmetry.

\begin{figure}[htbp]
\begin{center}
\includegraphics[width=9cm,keepaspectratio]{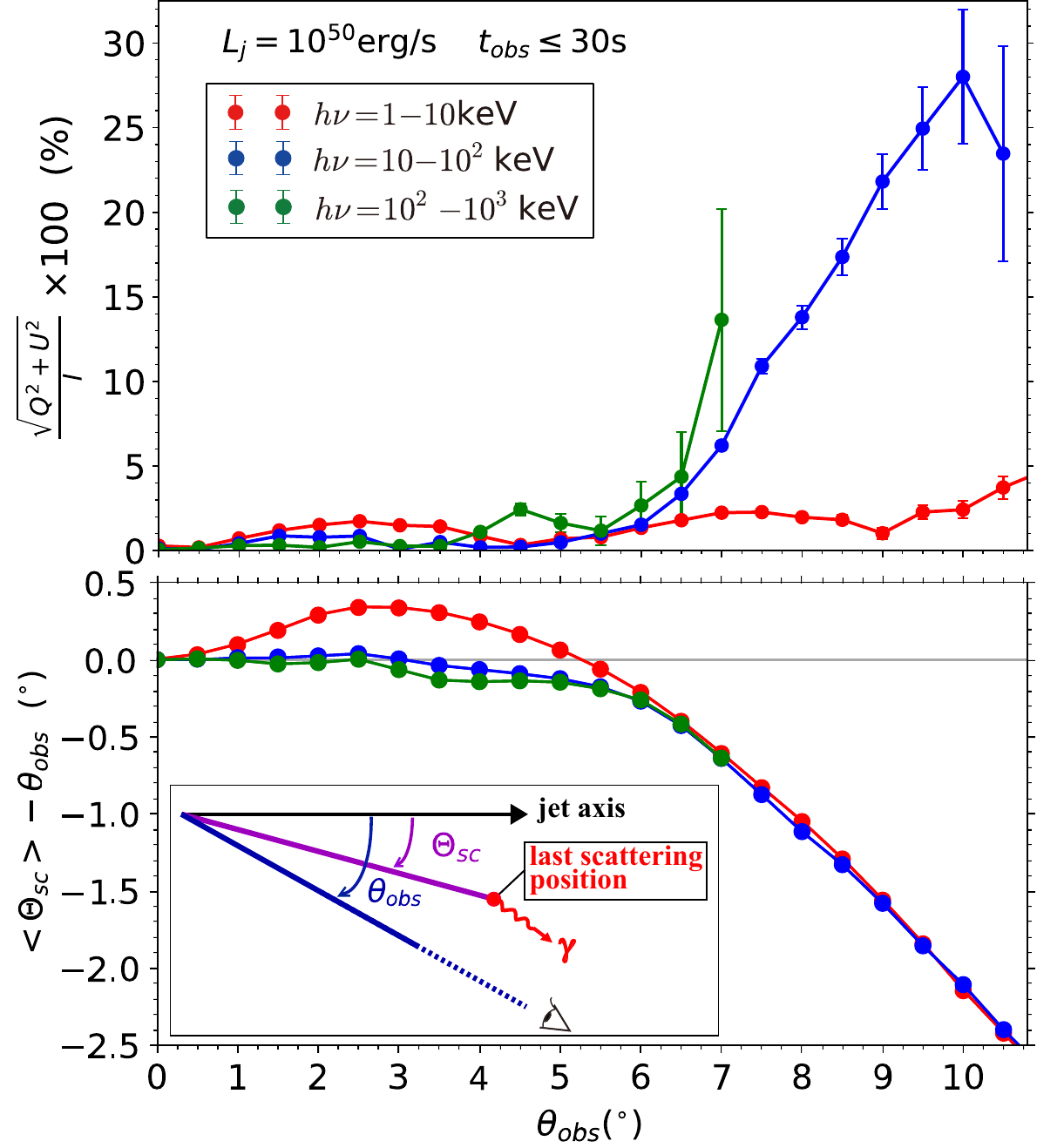}
\end{center}
\caption{Degree of linear polarization ($\Pi = \sqrt{Q^2 + U^2}/I$: {\it upper panel}) and the offset of an energy-weighted average value of the zenith angle of the last scattering position with respect to the viewing angle  ($\langle  \Theta_{sc} \rangle - \theta_{obs}$: {\it lower panel}) as a function of viewing angle for the model of $L_{\rm j}=10^{50}~{\rm erg/s}$ at $t_{obs} \leq 30~{\rm s}$. As in Figure \ref{fig:PDPAdist}, the color represents the different energy ranges, and the error bar in the upper panel indicates the  $1\sigma$ statistical uncertainty. For reference, the grey horizontal line in the lower panel shows the case in which the offset angle is zero ($\langle  \Theta_{sc} \rangle - \theta_{obs}$ = 0). The inset in the bottom panel provides a schematic picture describing how $\Theta_{sc}$ and $\theta_{obs}$ are defined.}
\label{fig:PDoffset}
\end{figure}

In Figure \ref{fig:PDoffset}, we demonstrate how the  anisotropy of the emission around the LOS is linked with the degree of polarization $\Pi$ by considering the model of $L_j=10^{50}~{\rm erg/s}$ with $t_{obs,max}=30~{\rm s}$ as an illustrative case. The top panel of the figure shows $\Pi$ as a function of $\theta_{obs}$, which is identical to the right half part of the central panel of Figure \ref{fig:PDPAdist}. On the other hand, the bottom panel shows the difference between the energy-weighted average value of the zenith angle of the last scattering position, $\langle \Theta_{sc} \rangle$, and $\theta_{obs}$. If the emission is isotropic around the LOS, $\langle \Theta_{sc} \rangle  =\theta_{obs}$ holds. In contrast, the offset between $\langle \Theta_{sc} \rangle$  and $\theta_{obs}$ tends to enlarge as the anisotropy grows. Therefore, the offset of the two angles,  $\langle \Theta_{sc} \rangle - \theta_{obs}$, can be regarded as an indicator of the anisotropy. As shown in the figure, the offset $\langle \Theta_{sc} \rangle - \theta_{obs}$ is modest at the core part of the jet ($\theta_{obs} \lesssim 3^\circ$) since the emission is close to symmetric around the LOS. Note that the slight offset of the average last scattering position toward a larger angle ($\langle \Theta_{sc} \rangle - \theta_{obs} > 0$) seen at the low energy band ($1$-$10~{\rm keV}$) reflects the fact that the outer part of the jet tends to show softer emission (i.e., intensity around the LOS at low energies ($h\nu < E_p$) tends to increase at larger $\Theta_{sc}$). Looking at larger viewing angles $\theta_{obs} \gtrsim 3^\circ$, the offset of the average last scattering position toward the smaller  angle ($\langle \Theta_{sc} \rangle - \theta_{obs} < 0$) is seen and tends to increase with $\theta_{obs}$. This development of  anisotropy arises due to the sharp negative gradient of the intensity in the lateral direction at the wing region. The link between the anisotropy of the emission and the degree of polarization $\Pi$ can be confirmed in the figure: $\Pi$ tends to increase as the offset between $\langle \Theta_{sc} \rangle$ and $\theta_{obs}$ becomes larger.

\begin{figure}[htbp]
\begin{center}
\includegraphics[width=9cm,keepaspectratio]{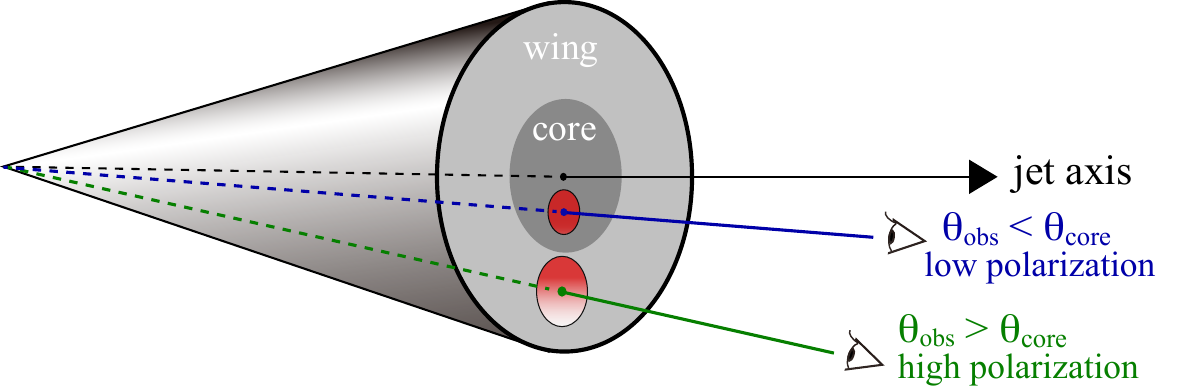}
\end{center}
\caption{Schematic picture of the intensity distribution for a given observer's LOS. In the picture, we show two cases: one in which the LOS lies within the jet core ($\theta_{obs}<\theta_{core}$: {\it blue}) and outside the core  ($\theta_{obs}>\theta_{core}$: {\it green}). The intensity around the given LOS is expressed by a color contour in red with a darker color corresponding to a larger intensity. As drawn in the picture, intensity is close to uniform at  $\theta_{obs}<\theta_{core}$, while a concentration toward a direction of the jet axis is seen at $\theta_{obs}>\theta_{core}$. Note that the picture is a simplified illustration of the overall trend, and the distribution of intensity in the simulation is more complex. For example, anisotropy in the azimuthal direction is also present in the simulation since the flow is not axisymmetric.}
\label{fig:PDoffset_Schematic}
\end{figure}

The above description of the anisotropy of the emission and the polarization is not limited to this illustrative case but is a general trend found in all cases explored in the current study. Hence, as mentioned earlier, while the viewing angle is within the core, a  low level of polarization is found,  and $\Pi$ tends to increase with $\theta_{obs}$ at a larger viewing angle. A schematic image for the anisotropy of emission around the LOS is displayed in Figure \ref{fig:PDoffset_Schematic}. The image illustrates how the distribution of the intensity changes with the viewing angle .

We find no general trend regarding the dependence of $\Pi$ on the energy band. When the viewing angle is small ($\theta_{obs} \lesssim \theta_{core}$),  lower energies ($h\nu = 1- 10~{\rm keV}$) tend to show the largest $\Pi$ due to the relatively enhanced anisotropy mentioned earlier. On the other hand, the highest energy ($h\nu = 100 - 1000~{\rm keV}$) tends to show the highest $\Pi$ as the viewing angle increases. This is because the intensity at higher energy ($h\nu > E_p$) shows a sharper negative gradient of intensity in the lateral direction. While $\Pi$ is in the range of a few $\%$ in most viewing angles, it can be larger than $10~\%$ in the wing region, particularly at high energies well above the peak ($\gg E_p$) and can be as high as $\sim 20 - 40\%$ in extreme cases. The non-monotonic energy dependence of $\Pi$ is similar to that found in \citet{IJT21}.

Comparing the results among the different choices of the integration time $t_{obs, max}$, the change in width of the core can be observed in the viewing angle dependence of $\Pi$. As described in Section \ref{sec:Jetstruc}, the core width increases at later times. Hence, the region that shows low polarization tends to extend to a larger $\theta_{obs}$ as the integration time increases. This trend is much more pronounced for the high-power jet model (see Section \ref{sec:Jetstruc} for the physical reason), as shown in Figure \ref{fig:PDPAdist}. We also find that, for a longer $t_{obs, max}$, $\Pi$ tends to be larger, particularly at the wing region. This is because the signature of the interaction between the jet and the stellar envelope becomes less significant at later times. As a result, the sharp velocity gradient structure at the wing becomes relatively more stable, which in turn leads to the enhancement of polarization.

It should be again noted, however, that the rapid expansion of the jet core may not take place for a higher spatial resolution simulation with a deeper injection radius (see Section \ref{sec:Ep-L}). A more conservative treatment would be to consider only the emission before the substantial expansion: $t_{obs} \lesssim 30~{\rm s}$ for $L_j = 10^{50}~{\rm erg/s}$   and  $t_{obs} \lesssim 10~{\rm s}$ for $L_j = 10^{51}~{\rm erg/s}$. 
Even if the affected emission times are excluded, our overall results remain valid: polarization is higher at large viewing angles because the velocity gradient is steep in the wings of the jet.

Looking at the position angle  $\psi = 1/2 ~{\rm arctan}(U/Q)$ in Figure~\ref{fig:PDPAdist}, a clustering around $\psi = 0^{\circ}$ ($180^{\circ}$) and  $\psi = 90^{\circ}$ is seen: the electric vector of the polarized emission is mostly inclined or perpendicular to the plane formed by the LOS and the jet axis. As mentioned above, $\psi$ is restricted to these two directions when axisymmetry is imposed. Hence, the result indicates that the departure from the axisymmetry in the current 3D simulations is not significant enough to wash out this feature. At the same time, however, a certain deviation from the axisymmetry is also confirmed by the intermediate values of $\psi$ found in the analysis. The departure from axisymmetry is more evident for the case of lower power jet model ($L_j = 10^{49}~{\rm erg/s}$) since jet dynamics  shows  wobbling behavior, which is induced during the jet-stellar interaction due to its lower thrust. It should also be noted that, as in the case of $\Pi$, the position angle shows a non-monotonic energy dependence. While $\psi$ can be aligned in all energy bands in some cases, the difference among the energy band can be as large as $~90^{\circ}$.
Such a feature is also indicated in \citet{IJT21}.  

\subsection{Time-resolved polarization}

\begin{figure*}[htbp]
\begin{center}
\includegraphics[width=17.5cm,keepaspectratio]{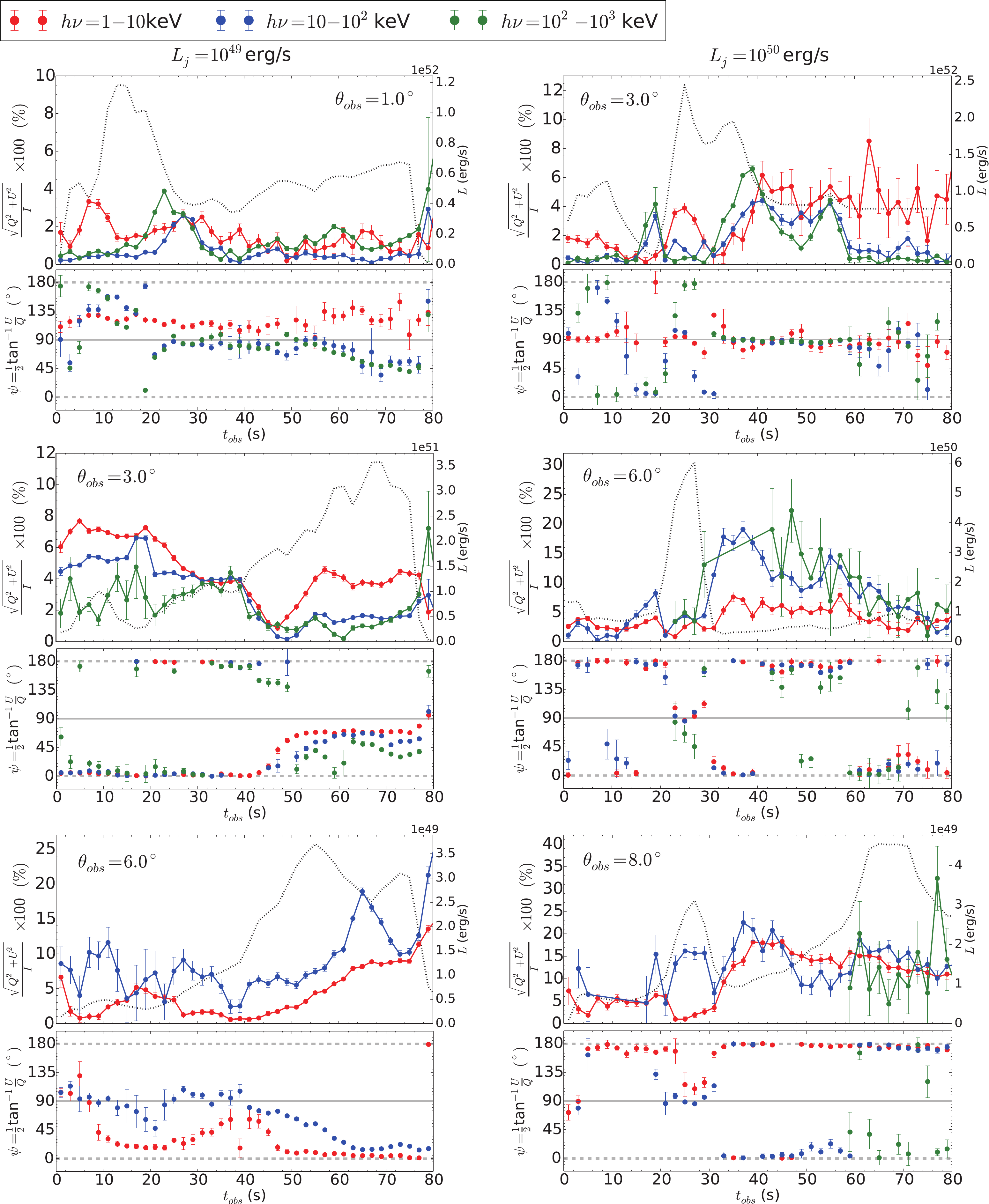}
\end{center}
\caption{Degree of linear polarization ($\Pi = \sqrt{Q^2 + U^2}/I$ : {\it upper panels}) and position angle ($\psi = 1/2 ~{\rm arctan}(U/Q)$ : {\it lower panels}) as function of observer time $t_{obs}$ at energy ranges of $h\nu = 1 - 10~{\rm keV}$ ({\it red}), $10-100~{\rm keV}$ ({\it blue}) and  $100-1000~{\rm keV}$ ({\it green}). The time bin employed for the analysis is $\Delta t = 2~{\rm s}$. The top, middle, and  bottom panels in the left (right) column correspond to the results of the model with jet power of $L_j=10^{49}~{\rm erg/s}$  ($10^{50}~{\rm erg/s}$) for viewing angles of $\theta_{obs} = 1^{\circ}$, $3^{\circ}$, and $6^{\circ}$ ($3^{\circ}$, $6^{\circ}$, and $8^{\circ}$), respectively. For reference, the corresponding light curve  is displayed by the black dotted lines.
The error bars represent the $1\sigma$ statistical uncertainty, which are evaluated using Equations (\ref{eq:pierr}) and (\ref{eq:psierr}).
We discard time bins where fewer than 100 packets were observed, due to the large statistical uncertainty. As for the position angle panels, we also omit points that show $1\sigma$ error bars larger than $40^\circ$. 
}
\label{fig:PDPAtevo}
\end{figure*}

The time-resolved analysis reveals a complex temporal behavior in the properties of polarization. As in the case of the light curves and spectra, these properties directly reflect the dynamical changes in the flow structure around the photosphere. In Figure \ref{fig:PDPAtevo}, we show the degree and position angle of the polarization as a function of time at different viewing angles for models with jet powers of $L_j = 10^{49}$ and $10^{50}~{\rm erg/s}$. The corresponding  light curves are also displayed in the figure for comparison. The figure confirms the presence of non-negligible time variability  in $\Pi$ and $\psi$. It also shows that there is no clear correlation between the temporal behavior of the polarization  with the light curves.

Similar to the results of the time-integrated analysis, the time-resolved values of $\Pi$ tend 
to increase with $\theta_{obs}$. The non-monotonic dependence of $\Pi$ on the energy band  found in the time-integrated analysis is also confirmed in the time-resolved analysis. It should be 
noted, however, some level of correlation is seen in  the temporal behavior of $\Pi$ among the 
different energy bands. For example, at relatively large viewing angles $\theta_{obs} \gtrsim 4^{\circ}$, $\Pi$ at all energy bands tends to increase at later times. As described in Section 
\ref{sec:PDPAtint}, this is because the jet-stellar interactions become less significant at 
later times.

Regarding the temporal evolution of $\psi$, it is concentrated  around the two orthogonal directions, $\psi = 0^{\circ}$ ($180^{\circ}$) and  $\psi = 90^{\circ}$, again indicating that the break of axisymmetry is not so significant. Simultaneously, though, our simulations show temporal change between the two orthogonal directions, i.e. a $\sim 90^{\circ}$ flip in $\psi$. Two examples of this behavior are visible in Figure~\ref{fig:PDPAtevo}: in the time period $t_{obs}\sim 20-30~{\rm s}$ for the  fiducial model ($L_{\rm j} = 10^{50}~{\rm erg/s}$), both the $\theta_{obs} = 6^{\circ}$ (middle right) and $\theta_{obs} = 8^{\circ}$ (bottom right) show such a flip. Within that period,  $\psi$ sharply changes from  $180^{\circ}$ ($0^{\circ}$) to $90^{\circ}$ and, after retaining its value for a few seconds, it again sharply changes back to $180^{\circ}$ ($0^{\circ}$). Interestingly, while the overall evolution does not suggest a correlation with the luminosity as mentioned above, the rapid rotation is associated with the spike in the light curve. As mentioned earlier, the deviation of axisymmetry is larger in the lower jet power model ($L_j = 10^{49}~{\rm erg/s}$). This is evident from the fact that the concentration of $\psi$ in the two orthogonal directions is less prominent in this model (left panels of Figure \ref{fig:PDPAtevo}). Accordingly, the rotation of $\psi$ also takes place relatively gradually compared to higher jet power models.

The overall temporal behavior of  $\psi$ at different energy bands appears to be somewhat correlated. Its value, however, can vary broadly, and the difference can be as significant as $90^{\circ}$ as already found in the time-integrated analysis. These results suggest that the non-negligible energy dependence of $\Pi$ and $\psi$ is an inherent feature of the photospheric emission. 

\subsection{Observational implications}

To date, observational data of GRB polarization is limited to a sample of relatively bright bursts. 
Therefore, to facilitate a comparison with these observations, it is necessary to concentrate on the emission emanating from the jet core with $E_p > 100~{\rm  keV}$, since this population is representative of the typical bright bursts.
As shown in the previous sections, the core region exhibits low polarization ($\Pi <$ a few $\%$). This result is consistent with the recent reports by the POLAR collaboration \citep{Kole20} and AstroSAT collaboration \citep{Chatto22}, which find that the majority of GRBs they have analyzed are consistent with having a low to null polarization across the full burst duration.
On the other hand, some of the bursts detected by AstroSAT and previous polarimetry missions \citep[e.g., GAP on board IKAROS;][]{YMG12} showed high levels of polarization, which cannot be explained by the emission emanating from the jet core region of our simulations.

Regarding the time-resolved properties, the large temporal change in the position angle ($\Delta \psi \sim 90^{\circ}$ flip) seen in the current simulation may be in line with the time-resolved analysis of polarizations in a few bursts \citep[GRB 110826A, 170114A, 160821A:][]{YMG11b, BKM19, Sharma19}. However, these analyses suggest  an intrinsically high degree of polarization is preferred in these GRBs, contrary to the intrinsically low polarization revealed in the current simulation. Accordingly, our simulations are also incompatible with the possible interpretation for the origin of low time-integrated polarizations found in the recent observations as a result of summation of intrinsically highly polarized emission that exhibits a large temporal change in the position angle \citep{BKM19, Kole20}. 

To sum up, our simulations predict a low intrinsic polarization for the emission originating from the core region, which is representative of the population of bright bursts.
 Therefore, while our results are in agreement with recent observations reporting low polarization for the majority of GRBs, they do not reproduce the high polarization inferred in some bursts.
It should be noted, however, that all observations so far accompany large error bars, and no robust consensus on the GRB polarimetry has been achieved.  Hence, we cannot draw a firm conclusion based on the available observational data.
%

\begin{figure*}[htbp]
\begin{center}
\includegraphics[width=18cm,keepaspectratio]{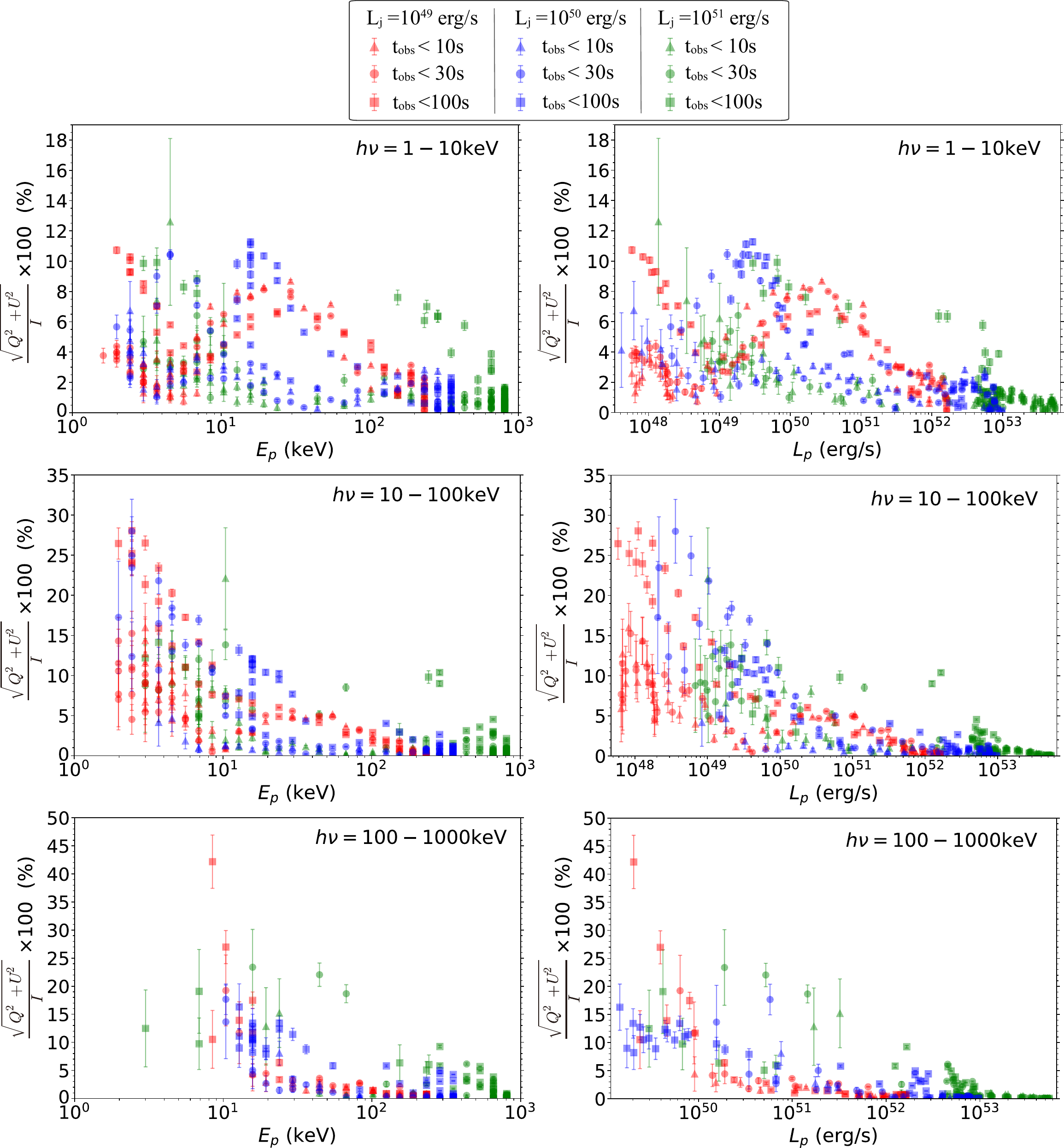}
\end{center}
\caption{Degree of linear polarization ($\Pi = \sqrt{Q^2 + U^2}/I$) as a function of spectral peak energy $E_p$ ({\it left}) and peak luminosity $L_p$ ({\it right}) at energy ranges of $h\nu = 1 - 10~{\rm keV}$ ({\it top}), $10-100~{\rm keV}$ ({\it middle}), and  $100-1000~{\rm keV}$ ({\it bottom}). 
Error bars indicate the $1\sigma$ statistical uncertainty evaluated as described in the caption of Figure \ref{fig:PDPAdist}.
  The color of symbols indicates the difference in the injected jet power: red, blue, and green correspond
  to $L_j = 10^{49}$, $10^{50}$, and $10^{51}~{\rm erg/s}$, respectively,
  The triangle, circle, and square symbols display cases in which emission up to a $t_{obs} = 10$, $30$, and $100~{\rm s}$ are taken into account, respectively. }
\label{fig:PDEpLp}
\end{figure*}

In the next decade, significantly more sensitive detectors dedicated to GRB polarimetry, POLAR-2 \citep{Kole19}  and LEAP \citep{McConnel21}, along with missions such as COSI \citep[][which, although not exclusively dedicated to GRBs]{Tomsick19}, are likely to provide further insights.
Here let us summarize the key characteristics of the polarization 
found in the current simulations, which might be tested in future missions. One of the robust 
features in the current results is the inverse correlation of the degree of polarization $\Pi$ 
with the spectral peak energy $E_p$ or the radiative output (peak luminosity $L_{\rm p}$ and 
isotropic energy $E_{iso}$). In other words, it predicts that dim, soft bursts  observed outside 
the jet core region ($E_p \lesssim 100~{\rm keV}$: XRR GRBs or XRFs) tend to show a higher 
polarization than the above-mentioned typical GRBs  viewed within the core ($E_p \gtrsim 100~
{\rm keV}$). To further examine this aspect, Figure~\ref{fig:PDEpLp} illustrates the 
relationship between $\Pi$, $E_p$, and $L_p$. Although the dispersion is large, the negative 
$\Pi - E_p$ ($L_p$) correlation can be confirmed from the figure. 


Another characteristic found in the time-resolved analysis is the clustering of $\psi$ in the two orthogonal directions associated with the rapid change between the two directions that occurs occasionally. The increased ability to perform a time-resolved polarization measurement in future missions may enable to probe these features.

Lastly, the non-negligible energy dependence of $\Pi$ and $\psi$ is also a notable feature found in the current study.  
Observations within the limited energy range of $\sim 10-1000~{\rm keV}$ alone by POLAR-2 or LEAP, may not be adequate for carrying out an energy-resolved analysis. 
 Hence, it is important to perform joint observations across different energy bands with other proposed missions, such as utilizing COSI and AMEGO at higher energies ($\sim 100~{\rm KeV}-5~{\rm MeV}$), along with instruments operating at lower energies like the Low Energy Polarimetry Detector (LPD) (see Section 7.1 of Gill et al. for details), in order to investigate the energy dependence of polarization.

\section{SUMMARY AND DISCUSSIONS}
\label{sec:summary}

In the present study, we have analyzed the properties of the photospheric emission in the context of LGRBs by conducting 3D relativistic hydrodynamic simulations and post-process Monte-Carlo radiation transfer calculations. This is an update of our previous work (IMN19) in which the number of photon packets tracked in the radiation transfer calculation was smaller by an order of magnitude. While we focused on light curves and time-integrated spectral properties in IMN19, the improved photon statistics here have enabled us to further explore the time-resolved spectral properties and the polarization properties.

We have considered three models in which the relativistic jet is injected at the interior of the massive stellar envelope with constant power ($L_j = 10^{49}$, $10^{50}$, and $10^{51}~{\rm erg/s}$) up to $100~{\rm s}$ with an opening angle $\theta_{ini} = 5^{\circ}$ (initially top-hat structure). Under the assumption that the electron distribution is Maxwellian, we have tracked the evolution of the photons subject to numerous electron scatterings before being released at the photosphere.
We properly take into account the full Klein-Nishina Compton section in determining the change of the Stokes parameter by the scattering. The summary of the main findings obtained from the current numerical models is listed below.

\begin{enumerate}
\item
  Due to the interaction with the massive stellar envelope, the jet develops a lateral structure in which the core region at the center ($\theta \lesssim \theta_{core}$)  is surrounded by a wing region that shows a rapid decline in power and Lorentz factor with angle (Figure \ref{fig:1}). 
  As a result, while the core region exhibits a spectral peak energy $E_p$ and luminosity $L$ close to uniform, these quantities sharply decline with the viewing angle for the emission arising from the wing region. This viewing angle dependence at $\theta_{obs} \gtrsim \theta_{core}$ leads to correlations between $E_p$ and the peak luminosity $L_p$ as well as between $E_p$ and the isotropic energy $E_{iso}$.
  The obtained $E_p$-$L_p$ correlation agrees well with the Yonetoku relation \citep{YMN04}, which further validates the key finding of IMN19 (left panel of Figure \ref{fig:EpLpEiso}). Additionally, we have demonstrated that the $E_p$-$E_{iso}$ correlation is broadly consistent with the Amati relation \citep{Amati02} (right panel of Figure \ref{fig:EpLpEiso}).

  

\item
  The viewing angle dependence also gives rise to correlations between the prompt emission properties ($E_p$, $L_p$ and $E_{iso}$) and bulk Lorentz factor of the flow $\Gamma_0$. The correlations are consistent with those inferred from observations \citep{Ghirlanda18} (Figure \ref{fig:EpLpGamma}).

\item
  The time-resolved spectral analysis reveals $E_p$-$L$ (luminosity) tracking feature within an individual burst emission (Figure \ref{fig:Lspec}), which is also indicated in the observations \citep{GMA83}. Moreover, the distribution of time-resolved $E_p$-$L$ data extracted from multiple bursts is fairly consistent with the Yonetoku ($E_p$-$L_p$) relation  as found in the study of \citet{LWL12} (Figure \ref{fig:E-Lcor}).

\item
  The time-resolved spectral fitting analysis shows that, while a thermal-like spectrum that can be fitted mostly by a cutoff power law (CPL) 
  function is found at the core region, the spectral shape tends to broaden with the viewing angle in the wing region (Figure \ref{fig:SPFIT}). As a result, while a fraction of low and high energy photon indices found in the analysis can be compatible with those of the observations, it is limited to the emissions arising from the wing region, which represents a population of soft dim bursts ($E_p < 100~{\rm keV}$: XRR GRBs and XRFs) (Figure \ref{fig:FITdata}). Hence, the current numerical models do not reproduce the observed non-thermal spectral features for the emission arising from the core region, which represents a population of the typical GRBs ($E_p \gtrsim 100~{\rm keV}$). 
  We discuss potential solutions to this problem later in this section.

\item
  The lateral structure of the jet is also imprinted in the properties of polarization. While the emission from the core region exhibits a low level of polarization ($\Pi \lesssim 4~\%$), the degree of polarization tends to increase with the viewing angle for the emission from the wing region and can be as high as $\Pi \sim 20$-$40~\%$, particularly at high energies ($h\nu > E_p$) (upper panels of Figure \ref{fig:PDPAdist}). Hence, the current simulations predict that the typical GRBs show lower polarization than their soft dim counterparts  (XRR GRBs and XRF) (Figure \ref{fig:PDEpLp}). 
  The low degree of polarization for the typical GRB is consistent with the latest result of POLAR \citep{Kole20} and  AstroSAT \citep{Chatto22} in which a majority of the observed bright bursts were found to be consistent with having a low to null polarization. 
  On the other hand, it is incompatible with a fraction of GRBs detected by AstroSAT and also with those detected by the earlier polarization mission \citep[e.g., GAP;][]{YMG12} 
  in which a high degree of polarization is indicated. However, all measurements have large uncertainty; therefore, we cannot draw a firm conclusion from the comparison.

\item
  While the current simulations are performed in 3D, the resulting jet structure does not show significant deviation from axisymmetry. 
  As a result, the position angle of the polarization $\psi$ tends to be clustered around the two orthogonal directions, which  correspond to the restricted angles for  axisymmetric outflow (lower panels of Figure \ref{fig:PDPAdist}).  The time-resolved analysis  finds a presence of $\Delta \psi \sim 90^{\circ}$ flip as a result of the transition between the two angles (lower panels of Figure \ref{fig:PDPAtevo}). This result may provide a possible origin for the large temporal variation of $\psi$ inferred in a few GRBs  \citep{YMG11b, BKM19, Sharma19}, although the intrinsic degree of polarization in the current models are much lower than those indicated in the observations.
   
\item
Both $\Pi$ and $\psi$ do not exhibit any clear correlation with the photon energy  (Figures \ref{fig:PDPAdist} and \ref{fig:PDPAtevo}). Although higher-energy photons tend to show larger $\Pi$, this is not always the case. Notably, it is interesting to observe that the difference in $\psi$ between different energy bands can reach up to $90$ degrees. Such properties offer an important test for the current numerical models, which can be examined in future polarization missions such as POLAR-2 \citep{Kole19} and LEAP \citep{McConnel21}.

\end{enumerate}

The current  and previous studies of photospheric emission  based on hydrodynamical simulations \citep[e.g.,][]{LMB13, PL18, PLL18, IMN19, IJT21} have shown that the emission mechanism provides a natural explanation for the origin of empirical correlations found in the prompt emission. The series of studies have also shown, on the other hand, that their spectral shapes turn out to be narrower than those of observed GRBs. The narrow spectral shapes are the main drawback in these global numerical models. As discussed in Section \ref{sec:spectral}, the tension may be attributed to the possible sub-photospheric dissipative processes that are not captured in the simulations.  Although implementing such effects in global simulations is numerically challenging, further investigation of this issue should be conducted to assess the role of photospheric emission in GRBs properly.

For the first time, we have quantified the GRB polarization signature based on 3D simulations. While the overall properties are similar to those found in the previous 2D simulations \citep{PLL20, IJT21, Parsotan22}, we also observe some differences due to the lack of axisymmetry (non-zero $\Pi$ at $\theta_{obs}=0^{\circ}$ and $\psi$ not being restricted to two orthogonal angles).
The characteristics revealed in the current study may be tested by future polarization missions,  which will provide more well-constrained polarization measurements. It should be noted, however, that the inclusion of subphotospheric dissipation mentioned above  may influence the resulting polarization properties \citep{LVB18}.

Lastly, let us comment on the  limitations  of the current study (beyond the absence of subphotospheric dissipation). One caveat is the spatial resolution.\footnote{Recently, \citet{Arita23} scrutinized the effects of spatial and temporal resolutions (defined by the number of snapshots used per unit time) in hydrodynamical simulations on radiation transfer calculation of MCRaT, a calculation method analogous to the one employed in our current study.
As for the temporal resolution,  we performed a convergence check using an approach similar to their study. In our simulations, we adopt a temporal resolution of 10 snapshots per second for the initial 300 seconds, which decreases to 1 snapshot per second thereafter. Through comparative analysis with results derived from lower temporal resolutions, we confirm that the radiation properties converge to a consistent solution at the current resolution setting.  Similar to the findings reported by \citet{Arita23}, our study also indicates that an insufficiently high temporal resolution can induce artificial Comptonization, which consequently hardens the high-energy spectral slope. As for the spatial resolution, we have not undertaken a convergence check. However, while the hydrodynamic properties have not reached convergence in the current setup, we do not anticipate the emergence of artificial Comptonization reported by \citet{Arita23} in our simulations. This expectation arises from our prescription for radiation transfer calculation: instead of assuming uniform physical quantities within each hydrodynamic mesh (as in their study), we employ linear spatial interpolation from neighboring mesh centers to determine local physical quantities. This approach circumvents abrupt changes in hydrodynamic properties across the mesh which gives rise to the artificial spectral changes. It is worth noting that our test calculations for a relativistic steady spherical outflow confirmed that our interpolation method accurately captures adiabatic cooling effects, even under the constraints of a low spatial resolution.
}
Since the  simulations cover a large spatial domain, small-scale structures are washed out due to the limited spatial resolution.
Consequently, the expected short time-scale variability in the light curve, which arises due to the jet-stellar interaction \citep[e.g.,][]{GLN19}, is largely smeared out.
Moreover, the jet injection radius employed in the current simulation may not be small enough to achieve numerical convergence, as discussed in Section \ref{sec:Ep-L} \citep[based on the result of][]{GLN19}. It is also highly desirable to expand the current models to explore the impact of central engine activity\footnote{The effect of intermittent central engine activity on the photospheric emission has been previously studied by \citet{PLL18} based on 2D hydrodynamical simulation.
It should be noted, however, that 3D effects are likely to play a crucial role in the intermittent jet \citep{GLN20}. } as well as the magnetic field \citep[the jet at its base can be unstable and be chopped for short period of time: ][]{2011NewA...16...46B}. Recent series of 3D hydrodynamical and magnetohydrodynamical simulations by \citet{GLN20, GBS20, GBL21} have demonstrated that these ingredients have an important effect on the dynamics and, therefore, on its resulting emission. Improvements to our numerical model to address the issues mentioned above will be the subject of  future studies.


\acknowledgments
 We thank B. Zhang, T. Parsotan, and P. Beniamini for fruitful discussions. This work was supported by JSPS KAKENHI Grant Number JP23H01178, 
 JP20K14473, 
 JP19K03878, and 
 JP19H00693 and 
 23-12-00220 of the Russian Science Foundation. Numerical computations and data analysis were carried out on Hokusai BigWaterfall system at RIKEN,   XC50  at Center for Computational Astrophysics, National Astronomical Observatory of Japan, and the Yukawa Institute Computer Facility. This work was supported in part by a RIKEN Interdisciplinary Theoretical \& Mathematical Science Program (iTHEMS) and a RIKEN pioneering project ``Evolution of Matter in the Universe (r-EMU)'' and ``Extreme precisions to Explore fundamental physics with Exotic particles (E3-Project)''.

\end{document}